\documentclass[10pt,journal,compsoc]{IEEEtran}
\ifCLASSOPTIONcompsoc
  \usepackage[nocompress]{cite}
\else
  \usepackage{cite}
\fi
\ifCLASSINFOpdf
  \usepackage[pdftex]{graphicx}
  \graphicspath{{images/}{icons-short-paper/}{inline-icons/}{./}} % where to search for the images
  
\else
\fi
\usepackage{hyperref}
\hypersetup{pdftex,
   pagebackref=true,
   pdfpagemode=UseNone,
   pdftitle={\@title},
   pdfauthor={\@author},
   pdfsubject={},
   pdfkeywords={},
   pageanchor=true,
   plainpages=false,       % for problems with page referencing
   hypertexnames=false,    % for handling subfigures correctly
   bookmarks=true,         % we want bookmarks
   bookmarksnumbered=false,% include the section numbers in the list
   bookmarksopen=true,     % In the list, display highest level only
   bookmarksopenlevel=3,   % display three levels of bookmarks
   pdfpagemode=UseNone,    % show just the page
   pdfstartview=Fit,       % defail page view is the whole page at once
   pdfborder={0 0 0},      % we don't want those silly boxes for links
   breaklinks=true,        % allow breaking links
   colorlinks=false,       % don't change coloring of links
   nesting=true,
   linktocpage}

\usepackage{microtype}

\usepackage{expdlist}
\usepackage{tikz}

\newcommand{\inlinevis}[3]{\raisebox{#1}[0pt][0pt]{\includegraphics[height=#2]{#3}}}
\newcommand{\tp}[1]{\includegraphics[width=\picturewidth]{#1}}

\newcommand{\icontext}{\texttt{Icon+Text}\,\inlinevis{-3pt}{1.3em}{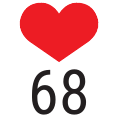}}

\newcommand{\textonly}{\texttt{Text Only}\,\inlinevis{-3pt}{1.3em}{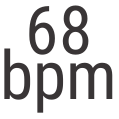}}

\usepackage[T1]{fontenc}
\usepackage{graphicx,calc}
\usepackage{multirow}
\newcommand{\bpstart}[1]{\vspace{1mm} \noindent{\textbf{#1.}}}

\newcommand{\health}[1]{\textcolor[rgb]{0,0.8,0.58824}{#1}}
\newcommand{\weather}[1]{\textcolor[rgb]{0.2,.6,1}{#1}}
\newcommand{\device}[1]{\textcolor[rgb]{1, 0.63137, 0.35294}{#1}}

\newlength{\picturewidth}

\newlength{\mytextsize}

\usepackage{tikz,pgfplots,pgfplotstable}
\usepackage{caption}  
\usepackage{makecell}
\usepackage{threeparttable}

\usepackage{soul}
\usepackage{xcolor,colortbl}

\usepackage{subcaption}
\usepackage{microtype}                 % use micro-typography (slightly more compact, better to read)
\PassOptionsToPackage{warn}{textcomp}  % to address font issues with \textrightarrow
\usepackage{textcomp}                  % use better special symbols
\usepackage{mathptmx}                  % use matching math font
\usepackage{times}                     % we use Times as the main font
\usepackage{cite}                      % needed to automatically sort the references
\usepackage{tabu}                      % only used for the table example
\usepackage{booktabs}                  % only used for the table example
\usepackage{cuted} %something for the appendix
\newcommand{\mc}[2]{\multicolumn{#1}{c}{#2}}
\definecolor{Gray}{gray}{0.85}
\definecolor{LightCyan}{rgb}{0.96,0.96,0.96}
\newlength{\maxlen}

   
\newcommand{\percentagebar}[1]{\begin{tikzpicture}[baseline = .15\mytextsize,every node/.style={inner sep=0,outer sep=0}]%
\draw[fill=white, thin,opacity=1] [yshift=1pt] (0,0) rectangle (\maxlen,.6\mytextsize);
\draw[fill=black, thin,opacity=0.7] [yshift=1pt] (0,0) rectangle (#1 \maxlen,.6\mytextsize);
\end{tikzpicture}
}

\usepackage{wrapfig}

\newcommand{\revision}[1]{\textcolor{black}{#1}}

\begin{document}

\title{Visualizing Information on Smartwatch Faces:\\
A Review and Design Space}

% \author{Alaul Islam, Tingying He, Anastasia Bezerianos, Tanja Blascheck, Bongshin Lee, and Petra Isenberg}
% \authorfooter{
% %% insert punctuation at end of each item
% \item
%  Alaul Islam, Tingying He, and Petra Isenberg are with Université Paris-Saclay, CNRS, Inria, LISN, Orsay, France. E-mail: {mohammad-alaul.islam,$|$\,tingying.he\,$|$\,petra.isenberg}@inria.fr.
% \item
% Anastasia Bezerianos is with Université Paris-Saclay, CNRS, Inria, LISN, Orsay, France. E-mail: anastasia.bezerianos@lri.fr.
% \item
% Tanja Blascheck is with University of Stuttgart, Stuttgart, Germany. E-mail: tanja.blascheck@vis.uni-stuttgart.de.
% \item
%  Bongshin Lee is with Microsoft Research, Redmond, USA. E-mail: bongshin@microsoft.com.
% }

\author{Alaul Islam, Tingying He, Anastasia Bezerianos, Tanja Blascheck, Bongshin Lee, and Petra Isenberg
\IEEEcompsocitemizethanks
  {
    \IEEEcompsocthanksitem 
    Alaul Islam, Tingying He, Anastasia Bezerianos, and Petra Isenberg are with the Universit\'e Paris Saclay, CNRS, Inria, France. E-mail: asif.alaul@gmail.com, tingying.he@inria.fr, anab@lri.fr, and petra.isenberg@inria.fr
      
    \IEEEcompsocthanksitem 
    Tanja Blascheck is with the University of Stuttgart, Stuttgart, Germany\protect\\ E-mail: research@blascheck.eu.
    
    \IEEEcompsocthanksitem 
    Bongshin Lee is with Microsoft Research, Redmond, WA, USA. E-mail: bongshin@microsoft.com.
}
	\thanks{}
}

% \thanks{Manuscript received January dd, 2023; revised January 26, 2023.}

%\markboth{IEEE Transactions on Visualization and Computer Graphics,~Vol.~X, No.~X, January~2023}%
\markboth{}%
{Islam \MakeLowercase{\textit{et al.}}: Visualizing Information on 
Smartwatch Faces}
% \markboth{IEEE TVCG Submission}%

\IEEEtitleabstractindextext{%
% \teaser
% {
%   % \centering
%       \includegraphics[width=\linewidth]{images/Teaser.pdf}
%     % \caption{}

\begin{abstract}
We present a systematic review and design space for visualizations on smartwatch\revision{es and the context in which these visualizations are displayed---smartwatch faces}. 
%We present a systematic review and design space for visualization on smartwatch faces.
A smartwatch face is the main smartwatch screen that wearers see when checking the time. Smartwatch faces are small data dashboards that present a variety of data to wearers in a compact form. Yet, the usage context and form factor of smartwatch faces pose unique design challenges for visualization. 
%We previously studied the display of data on smartwatch faces that were actively worn. Our goal was to provide information on wearers' interests and preferences to ground future work on smartwatch visualizations. 
%This paper builds on our prior work in multiple ways: 
%we conducted a second 
In this paper, we present an in-depth review and analysis of visualization designs for popular premium smartwatch faces based on their design styles, amount and types of data, as well as visualization styles and encodings they included. From our analysis we derive a design space to provide an overview of the important considerations for new data displays for smartwatch faces and other small displays. Our design space can also serve as inspiration for design choices and grounding of empirical work on smartwatch visualization design. We end with a research agenda that points to open opportunities in this nascent research direction. Supplementary material is available at: \url{https://osf.io/nwy2r/}.

\end{abstract}

}

\maketitle

\IEEEdisplaynontitleabstractindextext

% \fi
\IEEEpeerreviewmaketitle

\IEEEraisesectionheading{\section{Introduction}\label{sec:introduction}}
\IEEEPARstart{S}{martwatches} are powerful personal data collection devices and allow wearers to see various types of data measured from their body, their activities, or their environment. 
%Why is our work interesting?
Previous research has shown that people can perform simple comparison tasks with visualizations on smartwatches within several hundred milliseconds~\cite{Tanja:2019:Glanceable-Visualization}, providing evidence that visualizations are effective forms of data representations in the context of wearable devices. The usage context and form factor of small wearable displays, however, pose unique design challenges that require further dedicated research. 
%One can based decisions on encodings on empirical evidence, of which there is still little for smartwatch visualization, but also needs to consider the wearable characteristic of smartwatches.
In this work, we focus on reviewing watch face designs, \revision{the context in which visualizations are shown,} establishing a design space for smartwatch faces \revision{in terms of data and visualizations they can communicate}, and pointing to open research opportunities \revision{for smartwatch visualizations}. %, which are the primary screens~\cite{Gouveia:2016:Design-Space-Glanceable-Feedback} shown to wearers after a watch is turned on. 

Watch faces, the home screen of a watch, are the most frequently seen screen of a smartwatch. 
%The smartwatches typically display many different types of data to wearers at a glance.
One of the difficulties of visualization design for smartwatch faces is the many types of independent data (e.\,g., steps, weather, battery levels) that often need to be embedded in a coherent watch face design. These non-time/date functionalities on smartwatches\revision{, that can be represented as visualizations,} are called \emph{complications} in horology \cite{Jackson:2019:Design-Fundamentals}. Watch faces can be considered small personal dashboards with distinctive design challenges. The design challenges include a limited display space for possible complications, unique device form factors, the desire to express personality through watch face themes and design styles, and the mobile usage context that requires information to be readable at a glance. In addition, watch faces require that time or date is readable as the primary data. 

\begin{figure*}[t!]
\centering
\includegraphics[width=1\textwidth]{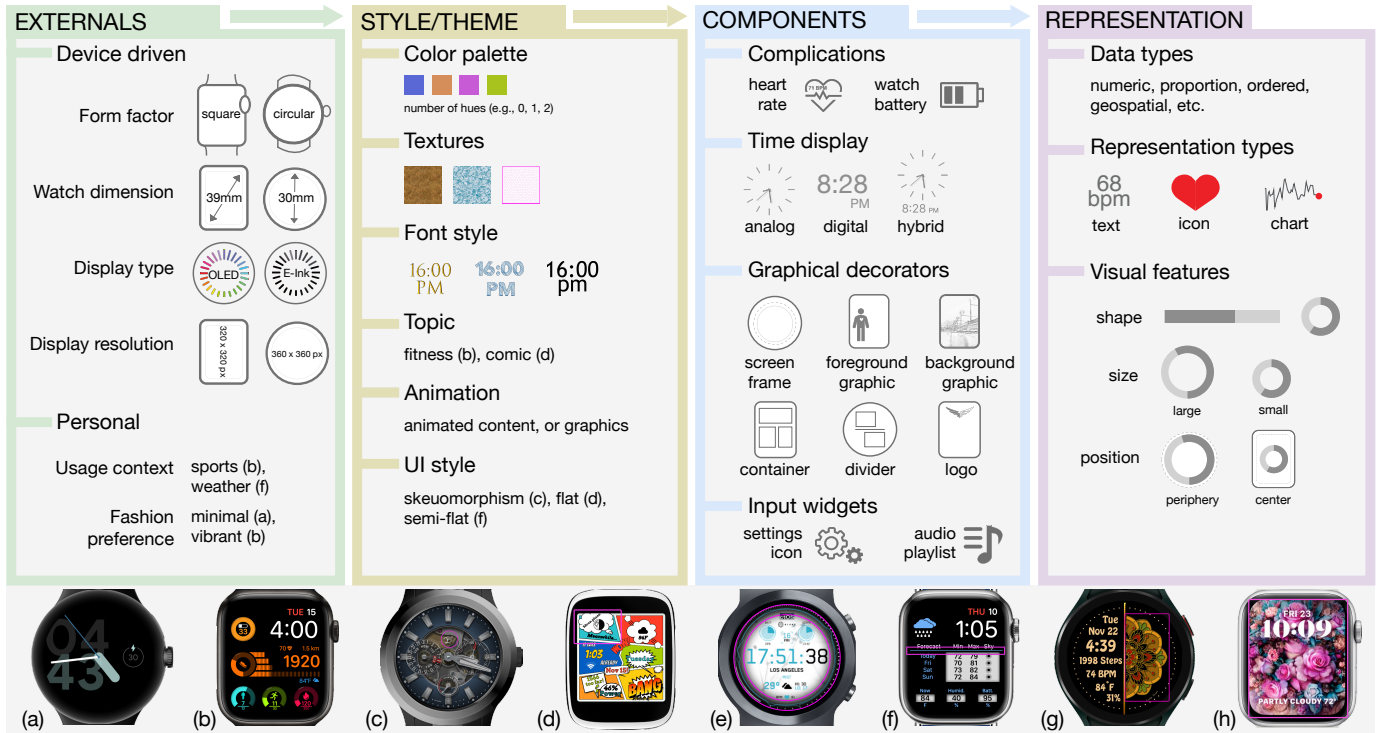}
 % \includegraphics[width=1\textwidth]{images/Teaser.pdf}
  % \includegraphics[width=.8\textwidth]{images/Teaser1.pdf}
  % \includegraphics[width=.8\textwidth]{images/Teaser2.pdf}
 % \vspace{-5mm}
  \caption{The dimensions of our smartwatch face design space \revision{that can influence visualization design and reading,} with examples. Watch face images (a, b, and f) indicate personal fashion preference as a factor; watch face images (c, d, and f) show examples of watch face design styles; and other watch faces (c to h) regions highlighted in magenta color are---a logo (c), a container (d), a screen frame (e), a divider (f), a foreground graphic (g), and a background graphic (h). \textit{Figure acknowledgements can be found before the references.}
  }
%\vspace{-3mm}
  \label{fig:teaser}
\end{figure*}

We conducted a first systematic investigation into the different constraints and considerations for the design of data representations for smartwatch faces. 
%
%Our work contributes a systematic analysis and a first design space, extending these past approaches. 
Specifically, we build on our prior short paper~\cite{Islam:2020:Smartwatch-Survey}, in which we report on a survey that asked smartwatch wearers to describe their current watch face and displayed data.
%of smartwatch faces showed that, on average, five different data items are displayed on wearers' watch faces and with some wearers reporting to see up to 17 items. In the survey, we investigated which data people consume and how it is visualized.
Our prior work, however, did not systematically analyze how visualizations were integrated in a coherent watch face: we did not investigate how charts were drawn, if watch faces used specific themes that applied to visualizations, or to what extent smartwatch face designers used graphical decorations \revision{that could affect visualization style and reading}. To address these limitations, in this work, we systematically analyzed 358 premium watch faces according to design dimensions, such as smartwatch components (e.\,g., time display, number of complications, graphical decorations) and visual features (e.\,g., interface styles, animations, color) \revision{as they could impact visualization style and reading}.
We compare our \revision{current review} to our \revision{prior survey} and derive a design space (see~\autoref{fig:teaser}) for visualizations on smartwatch faces. This design space reflects design considerations related to external and device factors,  watch face  themes and styles, watch face components, and their visual representation. We end with a research agenda and summarize future work on watch face visualizations. 

%with an in-depth analysis of premium smartwatch faces that wearers explicitly choose to use, a design space, and a research agenda that builds on our findings. %By targeting premium watch faces we concentrate our analysis on watch faces that wearers explicitly choose to use. %---because simply trying out watch faces would otherwise become a costly endeavor. 
%
%Limitations - keep for later
%We do not intend to give individual data design or smartwatch design concepts. Instead, we confer design features and limitations by investigating current smartwatch faces and provide recommendations for formulating grounded senses about possible smartwatch data representations.
%

%
%Our primary contributions are twofold: (i) a first in-depth description of the design space of smartwatch faces and their visualizations, and (ii) a research agenda to inform and inspire the visualization community to pursue and improve smartwatch visualizations.

% Smartwatch face examples (from~\cite{Facer:2020}). From left to right: a) Pixel-like Analog / Digital (clayton), b) Big step 6000 (MB-Watch), c) weather forecast (FAB Design), d) Comic Pro! (Round or Square) (Roch: Platinum Designs), e) B\# -- Open Heart (B Sharp Watches), f) Smart Flip Black (MR watchfaces\texttrademark), g) ADIDAS (Ddiego), h) MOD--486 Weather Watch (Michael O'Day), i)  Legion -- ctOS (Watch Dogs), j) WorldMap--Blue--S7 (Slumtek), k) Tokyo Tech -- Compressor (Tokyo Tech), l) TMJ | Active KL (TOMAJA Design).

\section{Related Work}
Most of past research on smartwatches has targeted smartwatches' technical capabilities (e.\,g., battery life~\cite{Min:2015:Battery-Use-Smartwatches}, resolution~\cite{Raghunath:2002:User-Interfaces-Wrist-Watch}, sensors~\cite{kamivsalic:2018:wrist-wearable-sensors-review}, interaction techniques~\cite{Neshati:2021:BezelGlide, Neshati:2022:EdgeSelect}, data collection methods~\cite{kim2022mymove, rey2023investigating}), or their role in people's life~\cite{Pizza:2016:Smartwatch-Vivo}. Much less work has focused on how to represent information on a small screen or how to design visualizations in the context of watch faces. The work most closely related to visualization design for smartwatches focused on particular design contexts such as health~\cite{Amini:2017:Data-Representations-Health, Suciu:2018:Active-Self-Tracking-VAS} or design ideations~\cite{carpendale:2021:Mobile-Visualization-Design-Ideation, Gouveia:2016:Design-Space-Glanceable-Feedback, Cibrian:2020:Children-ADHD-Design-Challenges} for smartwatch data representations. In this section, we first review work that solely discusses smartwatch visualization \revision{challenges} and then briefly summarize previous work on \revision{smartwatch representations on watch faces and in applications.}

{\subsection{Smartwatch Visualization Challenges}\label{subsec:SmartwatchVisChallenges}
\revision{\noindent \textbf{Tasks.} Albers et al.~\cite{Albers:2014:Task-Driven-Eval}  showed that tasks viewers conduct when exploring a visualization are influenced by the design of visual displays and choices of visual encodings, such as position and color. Additionally, the mapping variables employed in visualization, including the approach to data aggregation, also influence the viewers' tasks. For instance, a visualization may show the raw data or averages. Computed aggregates enable the visualization to perform tasks that would otherwise fall to the viewer.
% Additionally, the mapping variables employed in visualization, such as raw data or averages, and the approach to data aggregation also influence the viewers' tasks. 
Designing smartwatch faces to align with common tasks is important because designs may impact the wearers' tasks.
% These findings become particularly apparent when considering smartwatch face design, as it also has the potential to impact user tasks. 
The properties of smartwatch faces may also affect visualization reading tasks (e.\,g., space available, complications/visualizations shown), may constrain visualization design (overall style or theme that needs to be followed), and may be competing for visual attention (e.\,g., decorations). 
% These are all choices that a smartwatch face may influence: the space available, the position of the visualizations, the color schemes or styles the designer wishes to follow, etc. 
Smartwatches also present distinctive usage challenges that are much unlike the usage of visualizations on desktop screens. For example, on average people look at a smartwatch for just 5--7 seconds~\cite{Gouveia:2015:Habito,Pizza:2016:Smartwatch-Vivo,Visuri:2017:Smartwatch-Usage}, primarily to read the time (around 1.9 seconds~\cite{Pizza:2016:Smartwatch-Vivo}). The extent to which additional information can be understood from a quick look at the watch face remains uncertain. To answer this question, researchers studied low-level perceptual tasks to understand glanceability of smartwatch visualizations~\cite{Tanja:2019:Glanceable-Visualization, Blascheck:2023:Part-to-Whole-Glanceable-Vis}, the impact of visual parameters (e.\,g., size, frequency, color) on reaction times~\cite{Lyons:2016:Visual-Parameters-Smartwatches}, or representation preferences in an air traffic control usage scenario~\cite{Neis:2016:Feasibility-Airline-Crew-App}.}

\vspace{1mm} \noindent\revision{\textbf{Representations.}} \revision{Recently, researchers started to conduct dedicated research on dedicated visualization techniques for smartwatches. Some of these research efforts target
novel types of representations such as Chen’s temporal data~\cite{Chen:2017:Time-series}, Suciu
and Larsen’s time spiral~\cite{Suciu:2018:Active-Self-Tracking-VAS}, or Neshati et al.’s compressed line charts~\cite{Neshati:2019:G-Sparks, Neshati:2021:SF-LG}. In their exploratory studies, Amini et al.~\cite{Amini:2017:Data-Representations-Health} showed that minimal designs and simple data-driven visualizations in the form of charts have great potential to support in-situ data exploration on small smartwatch displays. Pekta{\c{s}} et al.~\cite{pektacs:2021:Smartwatch-Diabetes-Diary-Application} showed how simple diabetes-related visualizations using icons and emojis on warnings and alerts could motivate wearers to monitor health-related information. Gouveia et al.~\cite{Gouveia:2016:Design-Space-Glanceable-Feedback} proposed six design qualities for smartwatches through an iterative ideation process. These design qualities include abstraction, integrating with activities, supporting comparisons to targets and norms, being actionable, leading to checking habits, and facilitating engagement when designing representations for glanceable feedback on physical activity trackers.
}

\vspace{1mm} \noindent\revision{\textbf{Customization and Personalization.} Despite the growing capabilities of smartwatches, which can effectively capture and communicate a vast amount of information to wearers, some wearers risk abandoning these devices and missing out on potential benefits. This can occur when there is a lack of context-specific data representation or inadequate design of visualizations. Niess et al.~\cite{Niess:2020:Fitness-Tracker-Visualisations} studied the impact of various approaches to represent unmet fitness goals on trackers through visualization on rumination, highlighting that multicolored charts on fitness trackers may lead to demotivation and negative thought cycles. Havlucu et al.~\cite{Havlucu:2017:Tennis-Players} interviewed 20 professional tennis players and found that the abandonment of their fitness trackers was due to the type of information displayed on the smartwatches. The participants wanted specific tennis-related information, including recovery rate, nutrition, and details about their performance. They were interested in knowing where the ball hit their racket, the speed of a stroke, how the ball bounced off the floor, overall mobility on the court, and any weaknesses or errors in the game. Schiewe et al.~\cite{Schiewe:2020:SportsActivitiesonSmartwatches} studied real-time feedback during running activities for 40  participants. Their participants preferred visualizations over textual representations for self-serviced concurrent visual feedback on smartwatches.
}

\revision{
Our work focuses primarily on visualizations on smartwatch \emph{faces}. Being the home screen of a watch, watch faces are the most frequently seen screen of a smartwatch. It is important to understand more broadly how these small-screen displays are designed, as they provide the context in which visualizations may be seen. %Next, we discuss prior studies that discussed watch face data representations or application design. 
}

\subsection{Data Representations on Smartwatch Faces and in Smartwatch Applications}
% \subsection{Smartwatch Data Representations}
% We conducted a small systematic review to analyze papers on smartwatch face designs: \autoref{fig:smartwatch-related-publication} shows the review results. We collected papers using a snowball sampling technique that started from a first set of articles we had read for prior work. We complemented this approach with a Google scholar search with the terms: ``smartwatch visualization'' and ``smartwatch application,'' and looked at the first 550 results before they became largely irrelevant. We also searched the ACM and IEEE digital libraries with the term ``smartwatch'' and looked at the first 100 results. \review{After the first hundred initial results, we observed a repetition of works that had already been encountered during our earlier exploration through Google Scholar.} We only considered papers that discussed data displays on the smartwatch screen and excluded papers purely on technical considerations or interaction-focused content that did not discuss data displays.
% 46 
\revision{To review studies on data representations on smartwatch faces or in smartwatch applications, we collected publications that only discussed smartwatch data displays for applications, watch faces, or both.}  
% Among them, only 10 papers included watch face design ideas or prototypes. The other publications focused entirely on the design and development of smartwatch applications. We categorized the articles also according to their main research focus. 
\revision{Of the papers we found, five~\cite{Amini:2017:Data-Representations-Health, carpendale:2021:Mobile-Visualization-Design-Ideation, Esakia:2020:Smartwatch-Centered-Design, Klamka:2021:Bendable-Color-EPaper-Displays, Xu:2015:Shimmering-Smartwatches} focused on broad design ideations either on watch faces or on both watch faces and watch applications. Three papers~\cite{Blascheck:2023:Part-to-Whole-Glanceable-Vis, Islam:2022:Sleep-Data-Visualizations, Tanja:2019:Glanceable-Visualization} investigated the perception of data reading and comparison tasks on smartwatch visualizations.
The other papers focused on specific application contexts such as health~\cite{Neshati:2019:Displaying-Health-Data-Challenges, Suciu:2018:Active-Self-Tracking-VAS, Hansel:2016:Mood-Stress-Assessments-App, pektacs:2021:Smartwatch-Diabetes-Diary-Application}, enterprise applications~\cite{zenker:2019:collaborative-smartwatch-application, Christos:2018:Wearable-Cooperative-Assembly-Tasks-App, Bernaerts:2014:Office-Smartwatch-App, Bodin:2015:Security-Challenge-App}, driving~\cite{Lee:2016:Driver-Vigilance-Indicator-App, Li:2015:Driver-Drowsiness-Detection-App}, sports activities~\cite{Grioui:2021:Heart-Rate-Virtual-Smartwatch, Langer:2021:Mountain-Biking-Smart-Wearables, Asadi:2020:SmartWatch-based-Serious-Game}, map navigation~\cite{Wenig:2015:StripeMaps, Perebner:2019:smartwatch-pedestrian-navigation, Gedicke:2021:ZoomlessMaps}, child care~\cite{Cibrian:2020:Children-ADHD-Design-Challenges, Doan:2020:Children-ADHD-Application}, smart-home~\cite{Kosanovi:2018:SMART-HOME-App, Agnihotri:2017:Rist-Indoor-Navigation-App}, air traffic control~\cite{Neis:2016:Feasibility-Airline-Crew-App}, and library management~\cite{Wang:2017:Children-Search-for-Books-App}.
}

The works most closely related to ours concern smartwatch \emph{face} designs. We found only two papers~\cite{Gouveia:2016:Design-Space-Glanceable-Feedback, Esakia:2020:Smartwatch-Centered-Design} that focused entirely on smartwatch faces.
% , and eight others (see~\autoref{fig:teaser}) reported on both watch faces and applications. 
Gouveia et al.~\cite{Gouveia:2016:Design-Space-Glanceable-Feedback} focused on glanceable physical activity feedback for smartwatches derived through an iterative ideation process. They recommend glanceable feedback on the smartwatch face to prompt further engagement with the presented information. Esakia and Kotut~\cite{Esakia:2020:Smartwatch-Centered-Design} describe five guidelines for designing smartwatch applications and evaluated them in a mobile computing class, in which undergraduate students considered the applications as part of a design and development cycle. The design guidelines helped students develop watch faces in the context of a project promoting community physical activity.

With a similar focus on application areas, some research discussed representing health and physical activity data on smartwatch faces~\cite{Neshati:2021:SF-LG, pektacs:2021:Smartwatch-Diabetes-Diary-Application, van:2020:smartwatch-activity-coach, Neshati:2019:Displaying-Health-Data-Challenges, Suciu:2018:Active-Self-Tracking-VAS}. Recommendations for watch face designs from these papers were sparse. Neshati et al.~\cite{Neshati:2021:SF-LG}, for example, recommended a space-filling line chart technique and associated interaction techniques for time-series data. Van Rossum~\cite{van:2020:smartwatch-activity-coach} recommended aiming for understandable and clear visuals, especially for people with deteriorating eyesight, using a black background for contrast in dark environments. 

Others reported watch face design ideas from ideation activities, workshops, or design space explorations on smartwatch data representations~\cite{Klamka:2021:Bendable-Color-EPaper-Displays, carpendale:2021:Mobile-Visualization-Design-Ideation}. Carpendale et al.~\cite{carpendale:2021:Mobile-Visualization-Design-Ideation} offer new context-specific smartwatch applications and watch face design ideas from a range of situated ideation exercises. Klamka et al.~\cite{Klamka:2021:Bendable-Color-EPaper-Displays} show eight wearable application ideas and prototypes for personal information and mobile data visualizations with bendable color ePaper displays. In contrast, we focus on general smartwatch faces with a broad, everyday, usage context in mind. 

% As such, our work is most closely related to the following two papers. In their exploratory studies, Amini et al.~\cite{Amini:2017:Data-Representations-Health} show that minimal designs and simple data-driven visualizations in the form of charts have a great potential to support in-situ data exploration on small smartwatch displays. Gouveia et al.~\cite{Gouveia:2016:Design-Space-Glanceable-Feedback} proposed six design qualities: being abstract, integrating with activities, supporting comparisons to targets and norms, being actionable, having the capacity to lead to checking habits, and acting as a proxy to further engagement while designing glanceable feedback for physical activity trackers. 

% Our work complements these past recommendations by systematically reviewing commercially available smartwatch faces and structuring a visualization design space for smartwatch data to which such recommendations can be applied.

%In contrast to these works, our study contributes a review and discusses design considerations of watch faces that would benefit future research.

In our previous short paper \cite{Islam:2020:Smartwatch-Survey}, we studied which data people currently consume on their watch faces and how it is visualized. Our goal was to collect data that would allow to ground future studies on smartwatch visualizations (e.\,g., to inform choices about how many visualizations to test) in current practices of smartwatch wearers. Our prior work made the following contributions: we conducted an online survey with 237 smartwatch wearers, we conducted an online search of smartwatch face visualizations reprinted in \autoref{tab:exampleimages}, and we analyzed the potential for future smartwatch visualizations, in particular what data could be visualized by looking at the technical capabilities of the watches our participants wore.

Although this first survey yielded many interesting findings, the study had several limitations that we address in this article. Specifically, we did not collect (i) the types of visualization charts used on smartwatches; and (ii) how representations were integrated within other watch face visuals (such as themes). This work complements our prior survey by systematically reviewing commercially available smartwatch faces and structuring a visualization design space for smartwatch data.

\setlength{\picturewidth}{.115\columnwidth}
\newcolumntype{C}[1]{>{\centering\arraybackslash}m{#1}}

\begin{table}[t]
        \caption{Redrawn example representations from real smartwatch faces published in our short paper \cite{Islam:2020:Smartwatch-Survey}. Text colors correspond to three data categories: \health{Health \& Fitness}, \weather{Weather \& Planetary}, and \device{Device \& Location}. Bluetooth and wifi \emph{Only Text} and \emph{Only Icon} change color based on on/off status.}
    \renewcommand{\arraystretch}{1.2}    
    \centering
    \footnotesize
    \begin{tabular}{@{}p{.27\columnwidth}|@{}C{.13\columnwidth}@{}|@{}C{.13\columnwidth}@{}|@{}C{.13\columnwidth}@{}|@{}C{.13\columnwidth}@{}|@{}C{.13\columnwidth}}
    \toprule
        \textbf{Data Types} & \textbf{Only Text}&\textbf{Only Icon}&\textbf{Icon+ Text}&\textbf{Only Chart}&\textbf{Text+ Chart}\\
        \toprule
        \health{Heart rate / ECG waveform} & \tp{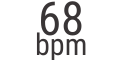}
        &\tp{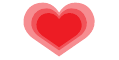}&\tp{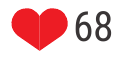}&\tp{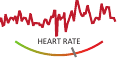}&\tp{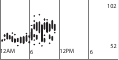}\\
        \health{Step count}&\tp{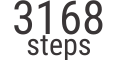}& &\tp{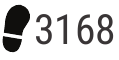}&\tp{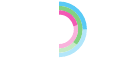}&\tp{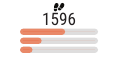}\\
        \health{Sleep related info}&\tp{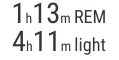}& &\tp{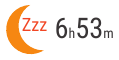}&\tp{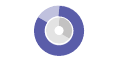}&\tp{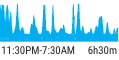}\\
        \health{Distance traveled}&\tp{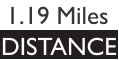}&&\tp{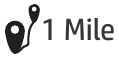}&\tp{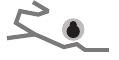}&\tp{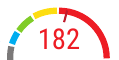}\\
        \health{Calories burned}&\tp{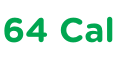}&&\tp{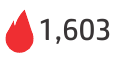}&\tp{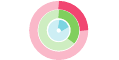}&\tp{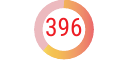}\\
        \health{Floors/Stairs climbed}&\tp{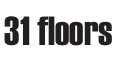}&&\tp{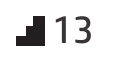}&\tp{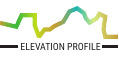}&\tp{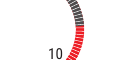}\\
        \health{Blood pressure}&\tp{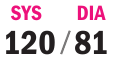}&&\tp{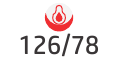}&&\tp{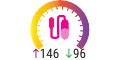}\\
        \hline
        \weather{Weather info}&\tp{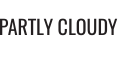}&\tp{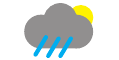}&\tp{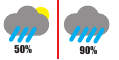}&&\tp{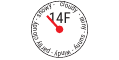}\\
        \weather{Wind speed/direction}&\tp{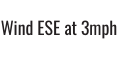}&&\tp{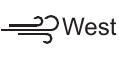}&&\tp{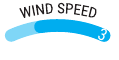}\\
        \weather{Temperature}&\tp{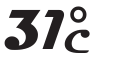}&&\tp{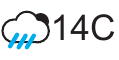}&&\tp{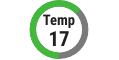}\\
        \weather{Sunset/Sunrise}&\tp{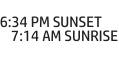}&\tp{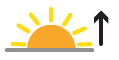}&\tp{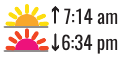}&&\tp{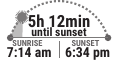}\\
        \weather{Moon phase}&\tp{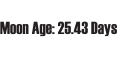}&\tp{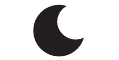}&\tp{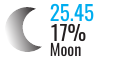}&\tp{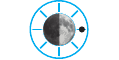}&\tp{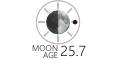}\\
        \weather{Humidity}&\tp{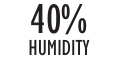}&&\tp{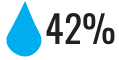}&&\tp{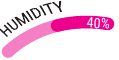}\\
        \hline
        \device{Bluetooth}&\tp{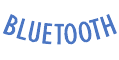}&\tp{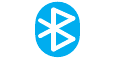}&&\tp{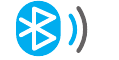}&\\
        \device{Phone battery level}&\tp{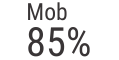}&&\tp{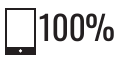}&\tp{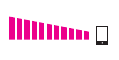}&\tp{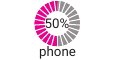}\\
        \device{Location name}&\tp{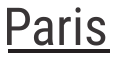}&&\tp{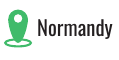}&\tp{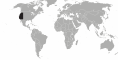}&\tp{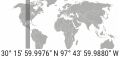}\\
        \device{Wifi}&\tp{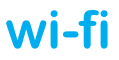}&\tp{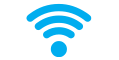}&\tp{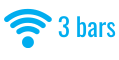}&\tp{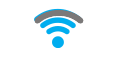}&\tp{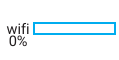}\\
        \device{Watch battery level}&\tp{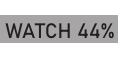}&&\tp{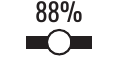}&\tp{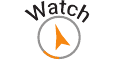}&\tp{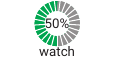}\\
    \bottomrule
    \end{tabular}
    %\vspace{-3mm}
    \label{tab:exampleimages}
\end{table}

\section{A Systematic Smartwatch Face Review}
To complement our prior survey~\cite{Islam:2020:Smartwatch-Survey} and to gain a deeper understanding of design considerations for smartwatch faces, we collected watch face designs and conducted a systematic review.
\revision{We study smartwatch faces because they are the context in which smartwatch visualizations are shown, and thus they may influence visualization design (e.\,g., if an overall style or theme needs to be followed) and visualization reading (e.\,g., available space, number of complications). Studying smartwatch faces, in particular data shown on them, also highlights opportunities for visualizations.}
In this section, we detail our methodology and report results. %Throughout this section, we refer to our previous work as \emph{the short paper survey} to distinguish it from the new and additional work, which we call \emph{the current review}.

\subsection{Data Collection and Analysis}
Our prior work \cite{Islam:2020:Smartwatch-Survey}  relied on participants' own descriptions of their smartwatch faces that we could not systematically verify. To address this limitation, we conducted a systematic review of premium (paid) watch faces from a popular watch face app and website. We focused on premium watch faces because we considered these to exhibit an acceptable level of design professionalism. %we considered 
%people's willingness to pay for these watch faces as an indication that they would actually use them on their smartwatch. 
The alternative of collecting watch face screenshots from a large number of participants would have been technically infeasible and potentially privacy invasive. Screenshots are difficult to take and transfer from watch to study platform for non-tech-savvy populations and watch faces often contain privacy sensitive information, such as locations or body measurements. 

We, therefore, decided to collect watch faces from the Facer App~\cite{Facer:2020}, one of the most popular smartwatch face distribution websites. It contains a Top100 page that lists the premium or free watch faces of Apple and watches running the WearOS/Tizen operating system. (Because WearOS and Tizen are closely related, we refer to both as WearOS from now on.) According to a Facer forum, the number of weekly syncs or downloads made decides the Top100 list. The list resets every Sunday at midnight (00:00). To become a premium designer on Facer, one must have three designs that received 3000 syncs in 30 days. 

% Because the list for the Apple Watch did not consistently contain 100 faces, we chose to focus on the WearOS/Samsung watch faces. 

We manually collected the metadata of the top 100 premium smartwatch faces every Sunday at midnight for one month, starting from March 14, 2021, for WearOS smartwatches and starting from September 18, 2022, for Apple smartwatches. When we began this work, the Apple list did not consistently contain 100 watch faces; therefore, we had to collect this data later. The metadata collected for each watch face included its rank, name, link, and thumbnail. 

Among the 800 top watch faces we collected, 358 were unique watch faces as several appeared in the top 100 for multiple weeks in a row. \revision{Some watch faces in the list were also extremely similar, and if so, we chose only the first instance collected during the initial week. To be considered unique, we looked at each watch face's distinctive characteristics: the data types shown, representation forms, UI style, time display, and graphical decorations (such as logos, backgrounds, or screen borders). For example, some watch faces differed only in a single complication or by color palette from the same watch face designer. We eventually selected 358 watch face designs from the initially collected 800 top watch faces.} 

To compare and analyze these 358 unique designs, we derived a set of codes that targeted an understanding of the overall design and how data representations were integrated in the watch face. We reviewed watch faces in the representative configuration shown on the Facer website (i.e., the configuration the watch face designer chose). Wearers can usually configure watch faces once installed, but we could not capture custom configurations. We acknowledge this as a limitation of our approach. \autoref{tab:top-dimensions} lists the codes we used. On a higher level, we focused the codes on \emph{what} is being represented and \emph{how} it is shown. We provide these codes as a supplementary file. %Next, we present our findings giving more information about the codes and, where appropriate, we contrast our findings to those of our prior survey~\cite{Islam:2020:Smartwatch-Survey}.

% (see~\autoref{Previous_Study}). Section 2.2 has been removed for the TVCG major revision 

We conducted all analyses using the extracted image of each watch face. In case a design was not clear from the thumbnail, we went to the Facer website to look at the simulated watch face graphic. We grouped our results according to the components shown on a watch face and how they were shown. We compared the results relating to the number and type of complications, time display, and complication representation between the current review and prior survey. We analyzed the results regarding graphical decorators, hues, animation, and UI style in the current review. 

\subsection{Results}
\subsubsection{What is Displayed on a Watch Face?}

\bpstart{Number of Complications} The watch faces contained a median of 4 complications similar to our prior work \cite{Islam:2020:Smartwatch-Survey}, in which participants reported a median of 5. However, we saw a difference in the number of watch faces with only one complication. While (13.41\% \percentagebar{.13}) (see \autoref{fig:complications_percentage}) of the premium watch faces contained only one complication, in our prior work \cite{Islam:2020:Smartwatch-Survey} only (0.84\% \percentagebar{.008}) of the watch faces contained one data item. The maximum number of complications per watch face drawn was similar: 16 in this review and 17 in the prior work \cite{Islam:2020:Smartwatch-Survey}.

\renewcommand\toprule{\specialrule{1pt}{1pt}{0pt}\rowcolor{gray!10}}
\renewcommand\midrule{\specialrule{0.4pt}{0pt}{0pt}}
\renewcommand\arraystretch{1.4}\tabcolsep3pt
\begin{table}[t]
    \caption{Codes used in the review of smartwatch faces.}
    \centering
    \label{tab:top-dimensions}
    \setlength{\tabcolsep}{3pt}
    \begin{tabular}{@{}p{.2\columnwidth}@{}p{.3\columnwidth}p{.47\columnwidth}@{}}
	\toprule
 \rowcolor{gray!10}  
	\mc{1}{}& Dimensions & \mc{1}{Values}\\
	\midrule
	\multirow{4}{4em}{Smartwatch Components} 
& time display & digital, analog, hybrid\\
& complication count & 0, 1, 2, \dots\\
& complication type  & watch battery, steps, calories, ... \\
& graphical decoration & container, logo, background/\-fore\-ground graphic, screen frame, divider \\
	\midrule
	\multirow{4}{4em}{Visual Features} 
& number of hues & 0, 1, 2, ...\\
&  UI style & skeuomorphism, semi-flat, flat\\
& animation & animation for graphical decoration, data-related animation (heart rate, watch battery, weather, etc.)\\
	\midrule
	Representa\-tion
	& type & text, icon, chart, or combinations\\
	\bottomrule
  \end{tabular}
\end{table}

\begin{figure}[tb]
\centering
  \includegraphics[width=1\columnwidth]{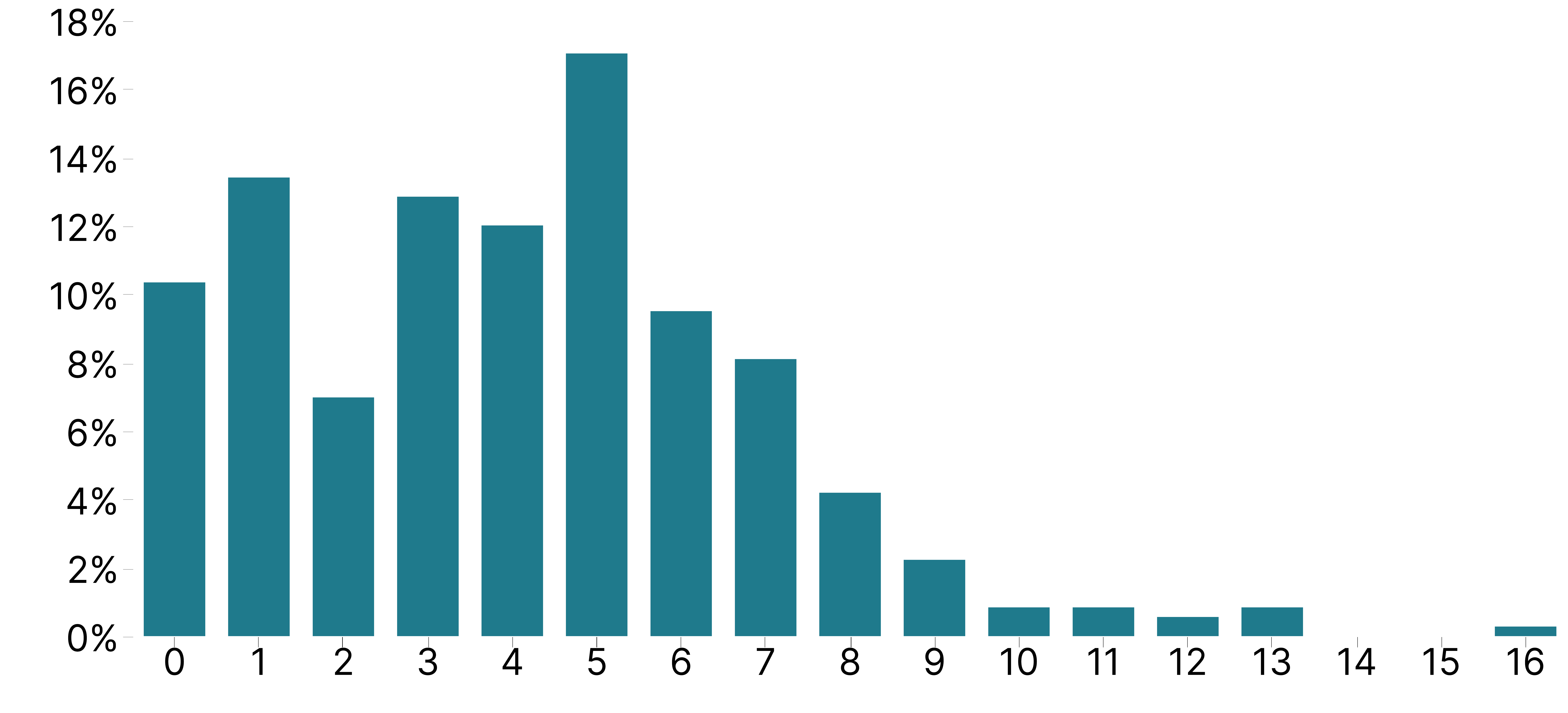}
 \vspace{-5mm}
  \caption{\revision{The percentage of watch faces (\emph{y}-axis) corresponding to different numbers of complications (\emph{x}-axis).}}
%\vspace{-3mm}
  \label{fig:complications_percentage}
\end{figure}

% \begin{figure}[tb]
% \centering
%   \includegraphics[width=1\columnwidth]{images/watchface_complications_comparison.pdf}
%  \vspace{-5mm}
%   \caption{Comparison of the percentage of respondents who saw a given number of complications in the survey and systematic review.}
% %\vspace{-3mm}
%   \label{fig:complications_percentage_comparison}
% \end{figure}

\bpstart{Types of Complications} \health{Health-fitness} data was the most common, followed by \weather{weather \& planetary} and \device{device-location} data echoing the results of our prior work \cite{Islam:2020:Smartwatch-Survey}. Watch battery level ranked first again as well. The main different to our prior work related to  Bluetooth connectivity data which was ranked 8\textsuperscript{th} previously \cite{Islam:2020:Smartwatch-Survey} but did not appear in the current review of data types, perhaps because Bluetooth icons only appear when a watch is paired to a phone and, therefore, did not show in our thumbnails.

% We compared the top 10 most common data types found in the systematic review to those from our survey (\autoref{fig:data_types_ranking}). Watch battery level ranked first both times. In both studies, \health{health-fitness} data were the most commonly reported, followed by \weather{weather \& planetary} and \device{device-location} data. Moon phase, exercise time, and standing time were in the top 10 most common data types in the review but not the survey (see \autoref{fig:data_types_ranking}). Bluetooth was ranked 8\textsuperscript{th} in the survey but did not appear in the review data types, perhaps because Bluetooth icons only appear when a watch is paired to a phone and, therefore, did not show in our thumbnails.

% \begin{figure}[tb]
% \centering
%   \includegraphics[width=1\columnwidth]{images/data_types_ranking.pdf}
%  \vspace{-5mm}
%   \caption{Ranking analysis of the most common data types from the survey (shown on the right) and found on the premium watch faces review (displayed on the left).}
% \vspace{-3mm}
%   \label{fig:data_types_ranking}
% \end{figure}

% Standing/movement time was ranked 12\textsuperscript{th} in the survey but ranked 10\textsuperscript{th} in the systematic review. Exercise/workout time was ranked 18\textsuperscript{th} in the survey but ranked 9\textsuperscript{th} in the systematic review. Moon phase was ranked 15\textsuperscript{th} in the survey but ranked 7\textsuperscript{th} in the systematic review. 

\subsubsection{How is the Watch Face Designed?}
Watch faces consist of components representing time, complications, and decorations, each with their own representation styles.

\bpstart{Time display}
\revision{We categorize watch faces based on the type of time display they employ, as different time displays may interfere with other visual components differently (e.\,g., analog hands may overlap visualizations and digital time displays may take up a considerable amount of space). Time displays can be divided into three categories: digital, analog, and hybrid, allowing us to categorize watch faces accordingly.}
% We divide watch faces into digital, analog, and hybrid watch faces depending on the time display. 
Digital watch faces represent time information as HH:MM:SS for hours, minutes, and potentially seconds. Analog watch faces typically use the hour, minute, and second hands to indicate the time, to resemble conventional analog watches. Hybrid watch faces have both digital and analog time displays. We found that the majority of premium watch faces were digital watch faces  (60.3\% \percentagebar{.60}), followed by analog watch faces (26.3\% \percentagebar{.26}) and last hybrid watch faces (13.4\% \percentagebar{.13}). The ranking of time displays was the same as in our prior work \cite{Islam:2020:Smartwatch-Survey} with small variations in the percentages (64.6\% digital, 23.2\% analog, 12.2\% hybrid). 

\bpstart{Complications Representations} We found seven ways complications were represented based on combinations of text, icons, and charts. As icons, we classified graphical content not in the strict semiotic sense and more analogously to how they are used in computing. Here, icons are a type of image that represents something else. As such, our icons can be both semiotic symbols \inlinevis{-2pt}{1.2em}{wifi-icon} and icons \inlinevis{-2pt}{1.2em}{weather-icon}. \autoref{fig:data_representation_observed} shows the representation types on average displayed on watch faces found in our review. \revision{The most common representation forms observed were text, icons, charts, and their combinations.} %, which closely align with the findings from our prior survey.}

\begin{figure}[tb]
\centering
  \includegraphics[width=1\columnwidth]{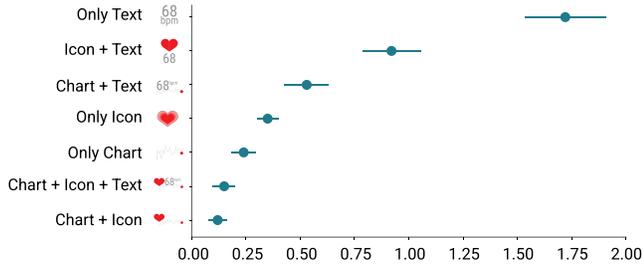}
 \vspace{-5mm}
  \caption{\revision{The average number (\emph{x}-axis) of representation types (\emph{y}-axis) presented on the premium watch faces.}}
%\vspace{-3mm}
  \label{fig:data_representation_observed}
\end{figure}

% \begin{figure}[tb]
% \centering
%   \includegraphics[width=1\columnwidth]{images/watchface_vis_comparison.pdf}
%  \vspace{-5mm}
%   \caption{\revision{The average number of representation types participants saw (survey) and presented on the premium watch faces (review).}}
% %\vspace{-3mm}
%   \label{fig:data_representation_observed}
% \end{figure}

A simple text label (\textonly) was the most common representation type for 1--2 data types on average on each watch face (\textit{M} = 1.72, 95\% CI: [1.53, 1.91]). Icons accompanied by text labels (\icontext) were the second most common (\textit{M} = 0.92, 95\% CI: [0.79, 1.07]). In our prior work \cite{Islam:2020:Smartwatch-Survey}, \texttt{Icon+Text} had been the most common representation type, used to display two kinds of data types on average on each watch face (\textit{M} = 2.05, 95\% CI: [1.78, 2.32]) followed by \texttt{Text Only} (\textit{M} = 1.38, 95\% CI: [1.13, 1.66]). Both results show that text seems to be the most frequent way to represent data on watch faces whiles charts or charts combined with text or icons are rare in practice. One notable difference in the data was the difference in \texttt{Only Icon} displays. Examples for representations that rely purely on a small image, such as weather icons (\inlinevis{-2pt}{1.2em}{weather-icon}\inlinevis{-2pt}{1.2em}{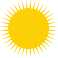}) are still rare on watch faces (see \autoref{tab:exampleimages}). Yet, in our prior work \cite{Islam:2020:Smartwatch-Survey}, surprisingly a large number of participants reported seeing \texttt{Only Icon} displays (\textit{M} = 1.11, 95\% CI: [0.93, 1.3]), which we attribute to a potential misunderstanding of the category. In our current review, \texttt{Only Icon} displays were, as expected, much more rare. We saw them for weather data (99\texttimes), moon phases (23\texttimes), wind directions (3\texttimes), compass/direction finder (1\texttimes), and altitude (1\texttimes).

%Although most of the time on watch faces, Icons are used for labeling or naming the data types, here in the review, we also observed how data could be shown with Icon Only. We discuss in more detail how data can be represented with the Icon Only in section \dots.

\vspace{6pt}
\noindent
For the following watch face components, we only report results from the current review because we had not asked participants about them in the prior work \cite{Islam:2020:Smartwatch-Survey}.

\bpstart{Graphical Decorators} We define graphical decorators as graphical content on the watch face that forms a coherent unit and takes up space, similar to complications, but does not carry any data. We found six types of graphical decorators~\revision{(see~\autoref{fig:graphical_decorators_example})}. \textit{containers}~(254\texttimes), which surround other watch face components; \textit{background graphics} (141\texttimes), which are decorations on the background such as a wallpaper; \textit{logos} (139\texttimes), which are small images to represent the brand or designer of the watch face; \textit{foreground graphics} (94\texttimes), which are other decorations in the foreground, such as small images or lines; \textit{screen frames} (48\texttimes), which are decorations on the border such as frames; and \textit{dividers} (31\texttimes), which are the lines that split the watch face into dedicated regions. 

\begin{figure}[tb]
\centering
  \includegraphics[width=1\columnwidth]{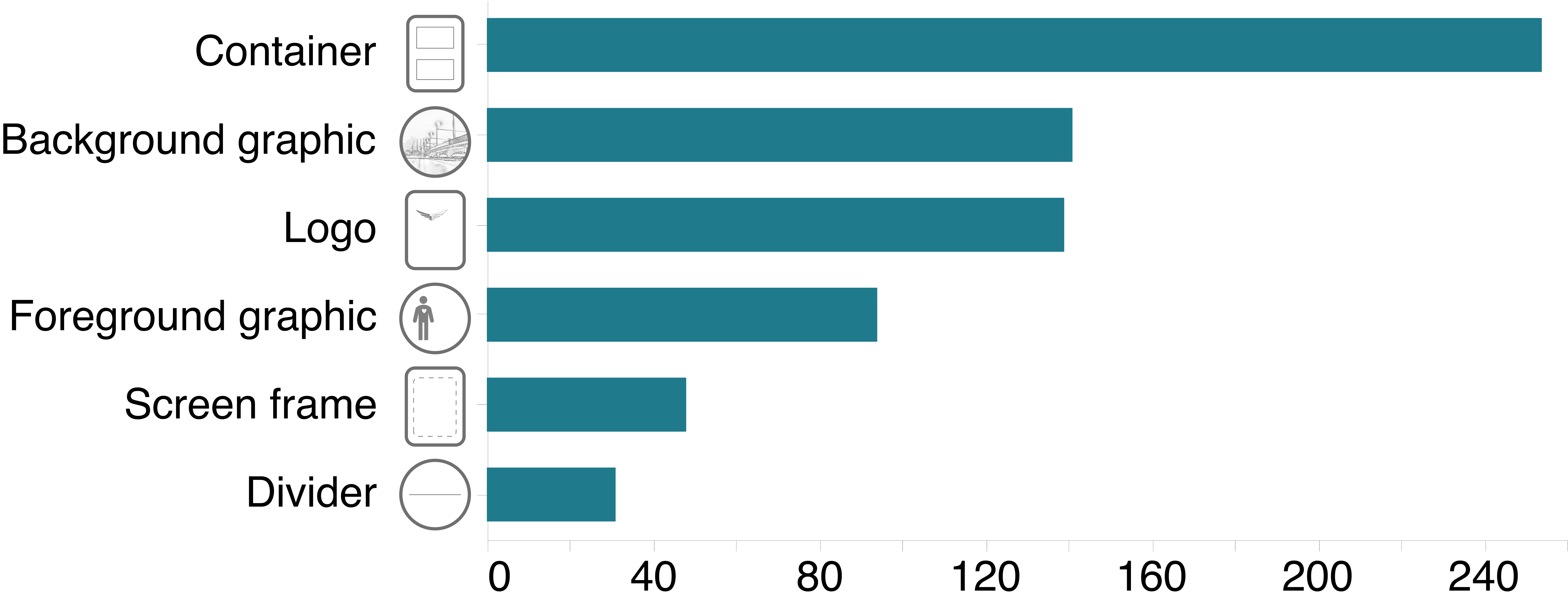}
 \vspace{-5mm}
  \caption{\revision{Number of watch faces (\emph{x}-axis) containing graphical decorators (\emph{y}-axis) found in the current review.}}
%\vspace{-3mm}
  \label{fig:graphical_decorators_example}
\end{figure}

% \begin{figure*}[tb]
% \centering
%   \includegraphics[width=2\columnwidth]{images/watchface_components_example.png}
%  \vspace{-3mm}
%   \caption{Graphical decorations on watch faces (from Facer~\cite{Facer:2020}). The region highlighted in magenta color is: (a) a container (Michael O'Day); (b) a logo (Ddiego); (c) a background graphic (Slumtek); (d) a screen border (Watch Dogs); (e) a foreground graphic (TOMAJA Design); (f) a divider (Tokyo Tech).}
% \vspace{-3mm}
%   \label{fig:graphical_decorators_example}
% \end{figure*}

\bpstart{Number of Hues} To get a better sense of the overall look of the watch faces, we analyzed the main hues used.  We did not consider black, white, and gray because these dominate most backgrounds. Nearly half of the premium watch faces (43.30\% \percentagebar{.43}) had only one hue (see~\autoref{fig:hue_percentage}). More than a quarter (24.58\% \percentagebar{.25}) had two, and only one-eighth (12.57\% \percentagebar{.13}) had three hues. The maximum number of hues on one watch face was seven.

\begin{figure}[tb]
\centering
  \includegraphics[width=.4\columnwidth]{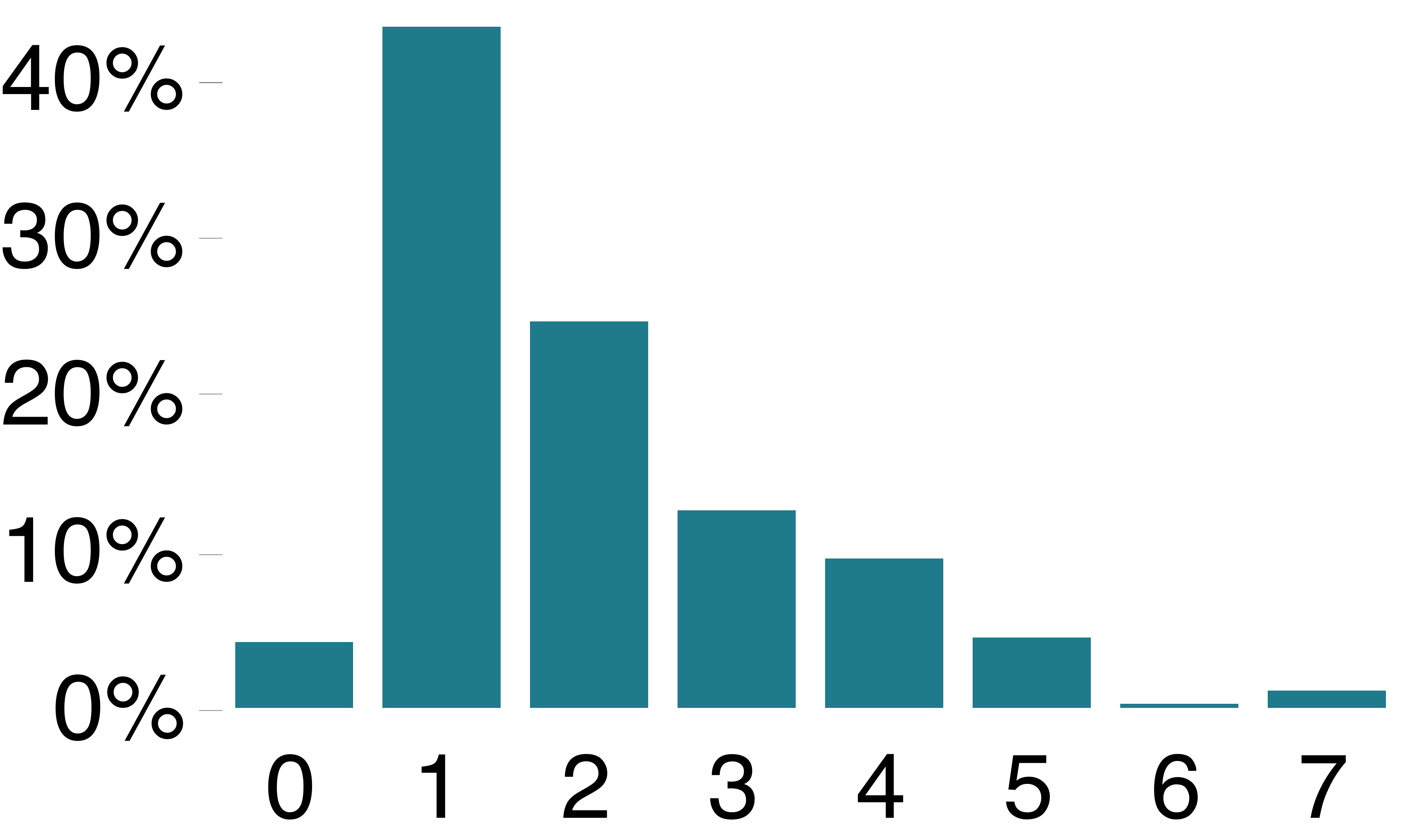}
 \vspace{-3mm}
  \caption{Percentage of watch faces (\emph{y}-axis) with a certain number of hues (\emph{x}-axis) in the current review.}
%\vspace{-3mm}
  \label{fig:hue_percentage}
\end{figure}

\bpstart{Animation} We found that 80\% (\percentagebar{0.80}) of the top watch faces had no animated content; 20\% \percentagebar{0.20} had animated content that did not represent data. Animations were used either on decorations or on complications to make them look more vivid and stand out. Sometimes animations were meant to make content even more iconic, such as a heartbeat animation for a heart icon\inlinevis{-2pt}{1.2em}{heart-icon}.

\bpstart{UI Style} Through the user interface (UI) style code, we captured the overall appearance of the watch faces. We found three categories of styles: skeuomorphism, flat designs, and semi-flat designs as shown in~\autoref{fig:teaser} (Style/Theme). In UI design, skeuomorphism is used to describe a graphical interface style, in which elements mimic their real-world counterparts~\cite{Skeuomorphism-definition:2022}. Skeuomorphism deploys gradients, shadows, or ornate details~\cite{spiliotopoulos:2018:ui-comparative} and on watch faces, these design elements are often used to recreate elaborate analog watch faces, such as those of pilot or diving watches. Flat design is a graphical interface style, in which no graphical elements attempt to create the appearances of continuous 3D depth. This style highlights simplicity by concentrating on two-dimensional elements, clean lines, and bright colors \cite{spiliotopoulos:2018:ui-comparative}. Semi-flat designs are in-between flat design and skeuomorphism: they are flat designs with some realistic touch, such as shadows. We found flat~(36.87\% \percentagebar{.37}) to be the most popular style, followed by skeuomorphism~(33.80\% \percentagebar{.34}) and semi-flat~(29.33\% \percentagebar{.29}) designs. The flat design was more frequent on Apple (56.32\%) than on WearOS watch faces (18.48\%), while Skeuomorphism was more frequent on WearOS (54.35\%) than on Apple watch faces (12.07\%).

\bpstart{Factors Influencing Design Popularity} \revision{Our metadata contained watch face ranks ranging from one to a hundred. We analyzed unique watch faces categorized by their ranks as coded in the metadata, aiming to uncover patterns and reasons behind the varying popularity of different designs. Watch faces ranked 1 to 50 exhibited an average of at least four complications, while there was a decline in complications beyond rank 50. Wearers seem to prefer watch face designs with a minimum of four types of complications, as supported by our prior work \cite{Islam:2020:Smartwatch-Survey}. We also explored representation types and found that trends in data representations were similar between top-ranked (1--50) and lower-ranked (51--100) watch faces. Minimal color usage was preferred in watch face designs, as evidenced by fewer hues in the top-ranked faces compared to those beyond rank 50. Additionally, we examined graphical decorators and found that in the top-ranked watch faces (1--50) graphical decorators were consistently more prevalent. These findings indicate that the number of complications, color usage, and graphical elements may play a role in shaping the preference for watch face designs among wearers. Future studies on design preferences will help gain deeper insights into what constitutes an appealing watch face design or a good design practice. A historical analysis in a few years might also reveal interesting results regarding preference changes and potential fashion trends.}

% \begin{figure}[tb]
%     \centering
%     \subcaptionbox{\label{fig:UI_flat}}
%     {\includegraphics[height=.30\columnwidth]{images/UI_flat.jpg}}
%     \subcaptionbox{\label{fig:UI_skeumorphism}}
%     {\includegraphics[height=.30\columnwidth]{images/UI_skeumorphism.jpg}}
%     \subcaptionbox{\label{fig:UI_semi_flat}}
%     {\includegraphics[height=.30\columnwidth]{images/UI_semi_flat.jpg}}
%     \caption{\revision{Smartwatch face user interface styles (from Facer~\cite{Facer:2020})---(a) Flat design (Dario Marnoni); (b) Skeuomorphism (B Sharp Watches); (c) Semi-flat design (MR watchfaces).}}
%     \label{fig:UI_style}
% \end{figure}
\section{Design Space}
Our review of existing watch face designs gave us the opportunity to reflect deeply on the design of watch faces. In this section, we generalize our findings into a broader design space. Our goal is to systematically  structure considerations for a holistic watch face, \revision{to shed light on the integration of visualizations}. 

Inspired by our previous evaluations, the design space is primarily descriptive \cite{Munzner:2022:DesignSpace}. It allows to describe existing watch faces based on a variety of factors. The design space also has the potential to be generative in that its individual components could inspire new watch face ideas. However, we have not yet verified the design space as such.
\autoref{fig:teaser} gives an overview of our design space with examples. The figure shows four main dimensions that are important to consider in the design of a watch face. 
The design space is based on our initial codes for the watch face designs listed in~\autoref{tab:top-dimensions}. Data collection and analysis of actual watch face properties led to the design space dimensions after several rounds of discussion among the co-author team. \revision{
The \emph{Style/Theme} dimension aligns with the \emph{Visual Features} from~\autoref{tab:top-dimensions}. To make the design space comprehensive, we included \emph{Font Style} and \emph{Texture} in this dimension, because our review and existing literature~\cite{Vanderdonckt:2019:Graphical-Adaptive-Menus} indicated their significant impact on the overall styling and theme of visualization designs. The dimension \emph{Components} is derived from the \emph{Smartwatch Components}, introducing \emph{Input Widgets}, which include interactive buttons/icons commonly found on watch faces but do not convey categorical or numerical data. The \emph{Representations} dimension emerged from the initial coding \emph{Representation Type}, encompassing \emph{Data Types} and \emph{Visual Features} influencing watch face visualization. We added an additional dimension in~\autoref{fig:teaser} called \emph{Externals}, which considers the smartwatch device itself and individual usage context, impacting the overall design. While not part of the watch face designs' coding or listed in~\autoref{tab:top-dimensions} these device and personal factors certainly may be important in the design of watch faces.
}

Unlike some other design spaces, our dimensions are not independent. They are ordered in a sequence (left to right) to show the dimensions' main direction of influence. We consider watch faces to be multi-view displays or data dashboards in which the designer carefully chooses what to include and how to style \& arrange them.
\revision{For developers taking inspiration from our design space, we recommend to complement their reading with a look into the specific developer guidelines for watch faces on the final platforms (Android~\cite{Android-developer-guides:2022}, Apple~\cite{Apple-developer-guides:2023}, etc.). These guides typically have additional broad guidance and feature descriptions available on each platform (such as automatic layouts, types of complications supported, etc.). Here, we cover platform-independent features needed to create holistic watch faces.}

% \autoref{fig:design_space} gives an overview of our design space. The diagram shows four main dimensions that are necessary to consider in the design of a watch face. In contrast to some other design spaces, our dimensions are not independent. They are ordered in a sequence to show the main direction of influence between these dimensions. We consider watch faces to be holistic multi-view displays or data dashboards in which the designer carefully chooses what to assemble, how to style it, and how to arrange it. However, it should be noted that the diagram is not necessarily a complete workflow diagram as certain decisions further down can also influence decisions higher up as part of a more lengthy and iterative design process. 

% \begin{figure}[tb]
% \centering
%   \includegraphics[width=1\columnwidth]{images/Smartwatch_Face_Design_Space.pdf}
%  \vspace{-3mm}
%   \caption{An overview of our design space. The diagram shows
% four main dimensions that are necessary to consider in the design of
% a watch face.}
% \vspace{-3mm}
%   \label{fig:design_space}
% \end{figure}

\subsection{Watch Face External Factors}
Factors external to the watch face may influence its design (\autoref{fig:teaser} Externals). The watch itself (device driven) defines the shape (form factor), display size (watch dimension), display type, and display resolution, which may influence choices in other parts of the design space. For example, the display type (such as OLED or E-ink) has a great influence on colors available to style the graphics and to represent data \revision{as visualizations}. 

E-ink watch faces, for example, have fewer colors and lower display refresh rates making animations difficult. They nevertheless have the potential for sport and activity smartwatches because of their paper-like characteristics, lightweight nature, low battery consumption, and  readability in sunlight~\cite{Angelini:2013:Smart-Bracelet}.
In contrast, OLED displays have greater color quality than other technologies. In recent years, OLED displays have risen to prominence as the main display technology for wearable devices. Most high-end smartwatches, including Apple and Samsung, use this technology. Due to the excellent image quality, OLED displays offer great possibilities for fashionable smartwatch face designs.

In addition to the display type, the display shape (form factor) has a profound impact on the layout and the graphics design. Watch faces now mostly come with square or circular displays; content is often shaped to match and create a harmonious aesthetic\revision{, which can, in turn, constrain the types of visualizations that can be shown}. Moreover, personal factors such as intended usage context or fashion preferences play an important role in the watch face design. Purchasing decisions may be made on whether a smartwatch would be wearable in casual or business contexts and whether it goes with a certain fashion, as seen in previous work~\cite{Schirra:2015:Uses-of-Consumer-Smart-Watches}. We expect that for some wearers similar decisions play a role for smartwatch faces and their design styles and overall look \revision{and consequently visualization use}.

\subsection{Styles, Themes, and Topics}
The theme of a watch face design describes a set of design attributes applied to all elements of the watch face to create a consistent, unified, or coherent look (\autoref{fig:teaser} Style/Theme). These attributes may define a set of colors used or visual textures imitated, the overall UI style, and potential animations used. Themes can also influence other dimensions such as the components of the watch and their representation. For example, a watch face with a minimalist theme may show only a few essential components such as a digital time, date, or battery life represented with simple visual features. 

A theme may relate to a topic such as fitness, outdoor sports, games, and fashion. For example, \autoref{fig:teaser}~(d) shows a dedicated comic theme with a flat UI style, a set of primary colors, and comic-style fonts chosen, while \autoref{fig:teaser}~(f) depicts a semi-flat UI style going for an appearance of vintage-style weather components. 

Watch face styles relate to themes and specify how specific components on the watch face are rendered. For example, in \autoref{fig:teaser}~(d), the fonts for labels use a comic style that matches the comic theme of the watch. 
Specific color-related design themes and styles are common on watch faces. For example, a darker color palette for the background is used most of the time perhaps to enhance content readability without disturbing wearers with bright light during the night or to save battery \cite{van:2020:smartwatch-activity-coach}. The flat and skeuomorphism UI style is extremely popular, whereas semi-flat designs are used frequently when the watch face is meant to be different and unique compared to traditional watches. Themes including animation often aim for a vivid overall feel but rarely use animation to represent data.

%
%Many watches do not have a precise topic and just aim for a minimalist style with a tasteful and aesthetic appearance based on a watch's core features. For example, a watch face with a clean and simple style may focus essential complications such as a digital time and date or battery life and might not require a lot of design considerations.
%However, a sport friendly themed smartwatch face considerably requires more design considerations. 

Both themes and styles influence how data representations on a watch face can be designed. Aiming for a consistent design would affect graphical embellishments or decorations such as shading used in visualization but also the available colors and potentially the types of visual channels that can be used. A watch face with a black-and-white theme, for example, cannot use multiple categorical color encodings based on hues. 

% \begin{figure}[tb]
%     \centering
%     \subcaptionbox{\label{fig:themecomic}}
%     {\includegraphics[width=.35\columnwidth]{images/ComicCrash-cut}}
%     \subcaptionbox{\label{fig:elegant}}
%     {\includegraphics[width=.35\columnwidth]{images/FreeGiaOpalLux-cut.png}}
%     \caption{Watch faces with a comic theme (a) Comic Crash by Roch Platinum Designs and an elegant theme (b) Free Gia Opal Cut by Gia (from Facer)}
%     \label{fig:themes}
% \end{figure}

\subsection{Watch Face Components}

We define components as any coherent object on the watch face that takes up display space, including complications, time-date-related data, graphical decorators, and input widgets (e.\,g., functional buttons) (\autoref{fig:teaser} Components). Designers have to decide which type of components to display and how many of them to put on the watch face. Only two types of components display data: time-related components (time, date, or day) and non-time-related components, called complications \cite{Android-developer-guides:2022} in horology. Graphical decorators refer to elements on smartwatch faces unrelated to data, for example, logos, screen frames, or background images. Input buttons such as settings or audio buttons are typically represented by icons on a touch-enabled smartwatch face and do not reflect any categorical or numerical data. Designers should avoid unfamiliar metaphors while designing or choosing such icons so that wearers can quickly understand their functions.
The choice of type and number of watch face components is closely related to the theme of the watch. For example, an aviation-themed  watch (\autoref{fig:teaser}~(c)) may require specific time-related components to show the time zone, a chronograph, and a date display, and it is likely to use a skeuomorphism design style. A fitness-themed watch face (\autoref{fig:teaser}~(b)) might instead require a large number of complications for showing step count, heart rate, or calories burned.  

The choice of an analog, digital, or hybrid display of time on the watch face also has an impact on the overall design. Analog watch faces have dials that may overlap complications and make them less readable. Digital watches use fonts that need to fit the general style and typically take up a prominent position on the watch face leaving less space for other complications.

Graphical decorators also play an important role in smartwatch faces. Through them, themes can often be expressed, and they, too, can constrain the space available for complications. Containers are a particularly frequent graphical decorator that partitions the watch face into regions and gives dedicated display space to complications. For example, \autoref{fig:teaser}~(d) uses a lot of space for comic-related background images and comic panel containers and constrains the complications to fit into speech bubbles. 
Decisions on which and how many components to display have an impact on how the components can be represented as discussed in the next section. 

% \textbf{Complications.} A complication is any feature in a watch face that is displayed in addition to time~\cite{Android-developer-guides:2022}. Complication types define what kinds of data can be shown in a complication. A watch face contains on average three to five complications. A space-efficient data representation technique on the watch face could be a designer's choice to meet the challenges. 

\subsection{Watch Face Data Representations}

Each watch face component's digital representation has at minimum a position, size, shape, and color scheme or texture (visual features). In this section, we focus our discussion on watch face complications that represent data other than time (\autoref{fig:teaser} Representation).

%To better understand how to represent data with one of the forms, we first discuss the smartwatch data in general without any classification (e.\,g., health data, weather data). Then we discuss the design considerations for the complications. 
%By the term “representation” on smartwatch face, we mean only the data representation or the data representation in complications.
One of the difficulties of designing a complication for smartwatches is that the data category it shows needs to be identifiable, for example, as steps, heart rate, or calories.  Designers can place signs to identify data categories or rely on wearers to learn and memorize a mapping. The types of signs that can identify a data category include---text labels that specify the category (e.\,g., ``steps''), accompanying icons (e.\,g., a foot icon for step count), and text units (e.\,g., ``km'' for distance traveled). The ``three rings'' on the Apple Watch is instead, an example of a mapping that requires learning \inlinevis{-2pt}{1.2em}{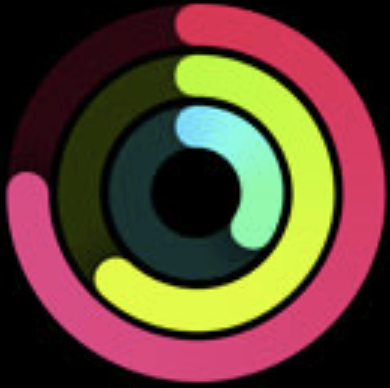}. It uses three concentric radial bar charts with memorable colors but without a sign to represent what each colored rings stands for: movement, exercise, and standing. One of the advantages of memorized mappings is that no display space needs to be dedicated to labels, units, or icons. 

In our evaluations, we observed that most smartwatches face complications represented data with combinations of text, icons, and charts. 
Most data encoded a single data value  
\inlinevis{-2pt}{1.2em}{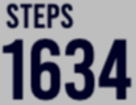}. More complex encodings, however, are certainly possible (e.\,g., multiple points over time or space \inlinevis{-2pt}{1.2em}{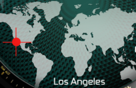}) and designers can take inspiration from more complex micro visualizations such as in work on word-scale visualizations \cite{Goffin:2017:AES,beck2017word} or data glyphs \cite{Fuchs:2016:ASR}. 

\subsubsection{Representations by Data Type}
We divided complications according to which data type they represent 
% as shown in \autoref{tab:common_data_abstraction} 
to discuss them in more detail: absolute numeric data, proportional data, categorical data, ordered data, temporal data, and geospatial data. Discussing complications by data type seemed more useful than by data category (e.\,g., steps, battery, \dots) as data in each category can easily be converted into different data types. For example,  a step count can be an absolute numerical value (e.\,g., 6700 steps), converted to a proportion (e.\,g., 67\% of a daily step goal), or shown as a time series. Next, we discuss representation possibilities and challenges based on the most common data types we saw represented on watch faces.

\textbf{Absolute numeric data} on watch faces commonly appears as either a single number or a pair of numbers. For example, the temperature can be a single numeric value (35°C) or pair of numeric values (Highest: 35°C; Lowest: 30°C).  Absolute numeric data is theoretically infinite or has a large range. This data can be represented as text or in a chart but their visual representation poses major challenges when the values dynamically update. There is often not enough space for visual references such as labels, tick marks, or grids in smartwatch complications. Without these references, one would have to predict data mappings that fit in the given display space and ensure the display adjusts to the dynamic updates without confusing wearers.
However, absolute numeric data often has a limitation in practice that can help to predict value ranges. For example, step counts can be unlimited. In reality, considering the time limit (e.\,g., within a day) and/or the limitation of human physical abilities, the value corresponding to the step count on the smartwatch face has a potential limit. It can help to consider plausible limits for reserving display space, calculating color scales, and identifying whether meaningful differences in values (e.\,g., 100 steps) should remain visible in a chart. So far, perhaps due to these challenges, charts that visualize absolute numeric data on smartwatch faces are still rare. \texttt{Only Text} representations are the more common representation form. 

\textbf{Proportion data} represents values according to a given maximum, typically 100\%. Visualizations of proportions are common on watch faces: in \autoref{tab:exampleimages}, most of the charts in the last two columns show a proportion. We found data in the form of real proportions and derived proportions. For real proportions, the shown value corresponding to this data category has originally been captured or measured as a proportion, such as 42\% humidity or  80\% power (of a fully charged battery). Derived proportions are proportions converted from absolute numeric data to a proportion according to a pre-set maximum and minimum. One of the common derived proportions is from a goal watch wearers set, such as steps---67\% (of 10k steps), calories burned---67\% (of 3~kCal).

\textbf{Categorical data} is a type of data that does not have an implicit ordering. Categories only distinguish whether two things are the same or different \cite{munzner:2014:visualization}. Categorical data is often represented by small graphical elements. Common are icons that stand for the categorical value (weather---rainy  \inlinevis{-2pt}{1.2em}{weather-icon}/sunny \inlinevis{-2pt}{1.2em}{sun-icon}) or a sign with an indexical color (Bluetooth---on \inlinevis{-2pt}{1.1em}{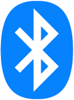}/off \inlinevis{-2pt}{1.1em}{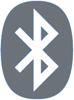}). 

As opposed to categorical data, \textbf{ordered data} does have an implicit ordering \cite{munzner:2014:visualization} (e.\,g., a low/middle/high level of the smartwatch battery). Most ordered data on watch faces is quantitative or has been derived from quantitative data. Icons or charts commonly represent ordered data with an ordered color scale encoding. Often, derived ordered data is encoded together with proportions. For example, battery indicators often turn red when the battery is almost depleted and green when it is fully charged. 

\textbf{Temporal data} on smartwatch faces most commonly relates to specific time points such as sunrise or sunset time, time when one went to sleep or woke up, etc.  
\begin{wrapfigure}{r}{1.5cm}
\centering
    \includegraphics[width=1.5cm]{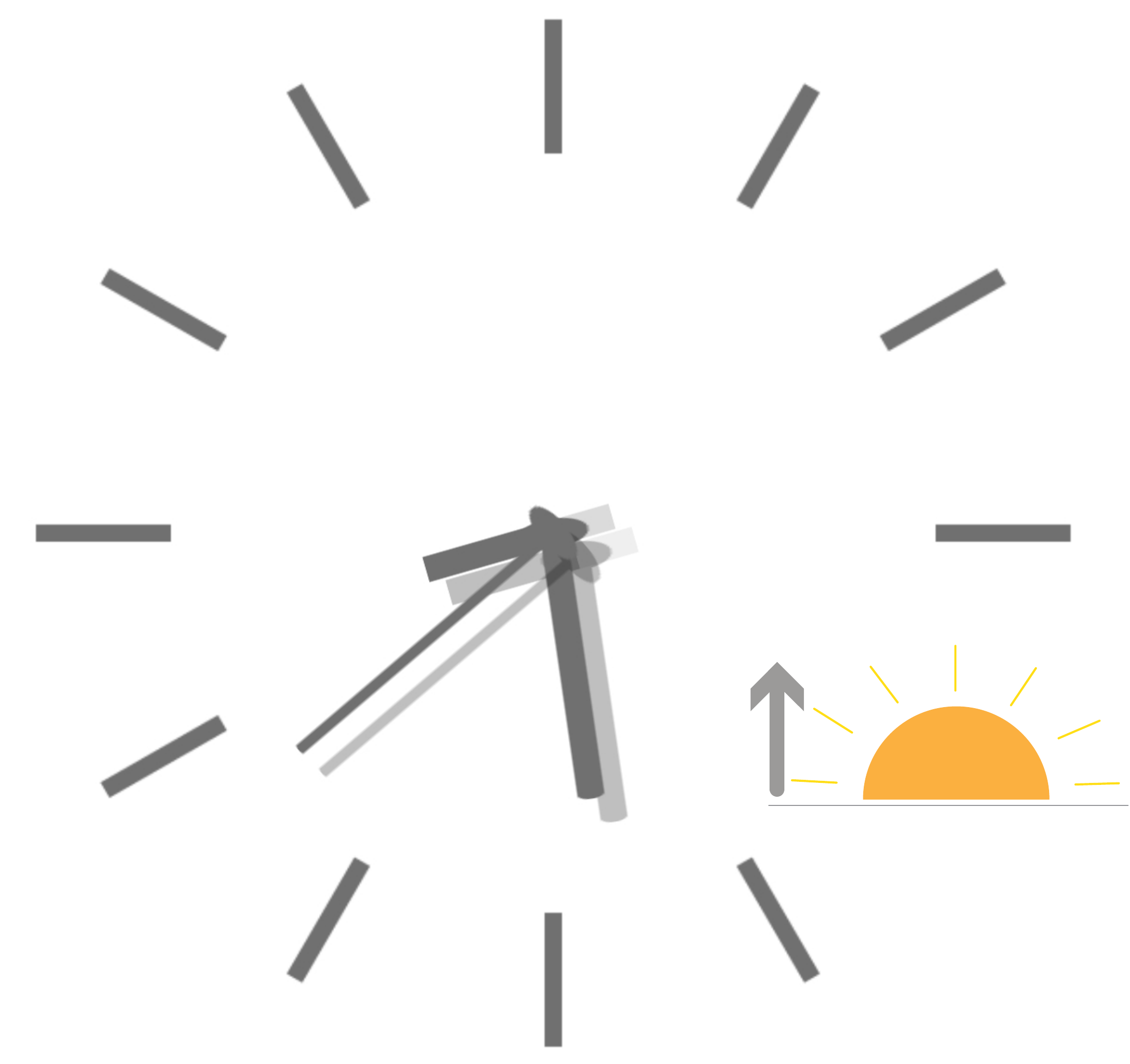}
    	\caption{A watch face showing sunrise time.}
     \label{fig:sunrise}
\end{wrapfigure}
Time series on a continuous scale are common for sleep data or heart rate.  Designers often visualize some temporal data in conjunction with the dial of analog watch faces, such as sunrise and sunset times (\autoref{fig:sunrise} shows an example of sunrise time) that are displayed as icons next to the hour numbers. Nevertheless, the variety of temporal visualizations on watch faces is still small compared to the many visualizations for time that exist \cite{Aigner:2011:TimeViz}. Especially designs for events and intervals are still rare on watch faces and how to port past ideas to a watch should be further explored (e.\,g., \cite{Dragicevic:2022:SpiraClock}). 

The most common \textbf{geospatial data} on smartwatch faces concerns the wearer's location and time zone.  In addition to text and icons (e.\,g., landmarks~\inlinevis{-2pt}{1.2em}{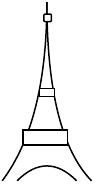}) maps~\inlinevis{-2pt}{1.2em}{inline-icons/05complicated_encoding} also sometimes represent this data. 

\subsubsection{Representation Types}
\textbf{Text} is the most common form of data representation on smartwatch faces. Text can be styled through font type, size, and color, which can also be used to encode additional information, for example, as in word clouds. However, we rarely saw font-size based encodings on actual watch faces despite them having been mentioned in research \cite{Amini:2017:Data-Representations-Health}.

\textbf{Icons} on watch faces are widely used for labeling but rarely to represent data. They can be used for categories if icons exist to meaningfully identify the categories. This is the case for weather data where viewers are familiar with icons that represent a rainy\inlinevis{-2pt}{1.2em}{weather-icon} or sunny \inlinevis{-2pt}{1.2em}{sun-icon} day. It would be harder to find icons that represent REM, Light, or Deep sleep stages. 

There are two options to use an icon to represent quantitative data: a) designers can turn the quantity into an order or category and use a visual variable for categories such as hue, texture, or position or b) add a visual variable to the icon that can represent a quantity such as position, size, area, flicker, etc.  For example, suppose a designer wants to represent the quantity of calories burned with an icon. In that case, they first have to choose an icon that represents calories; a fire icon \inlinevis{-2pt}{1.1em}{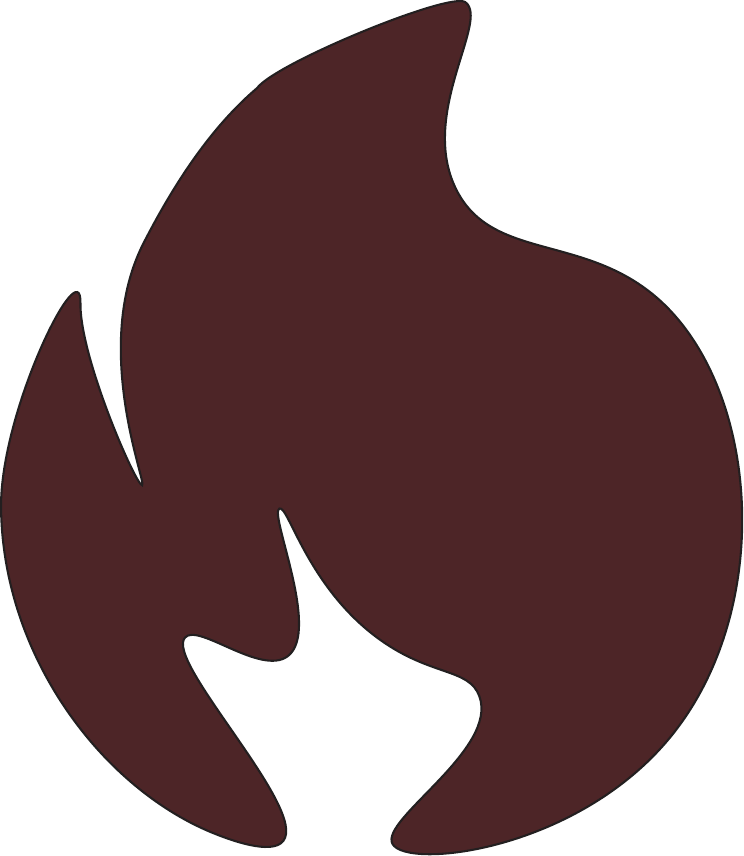} is relatively common. Next, the designer could apply a sequential color scale (\inlinevis{-2pt}{1.1em}{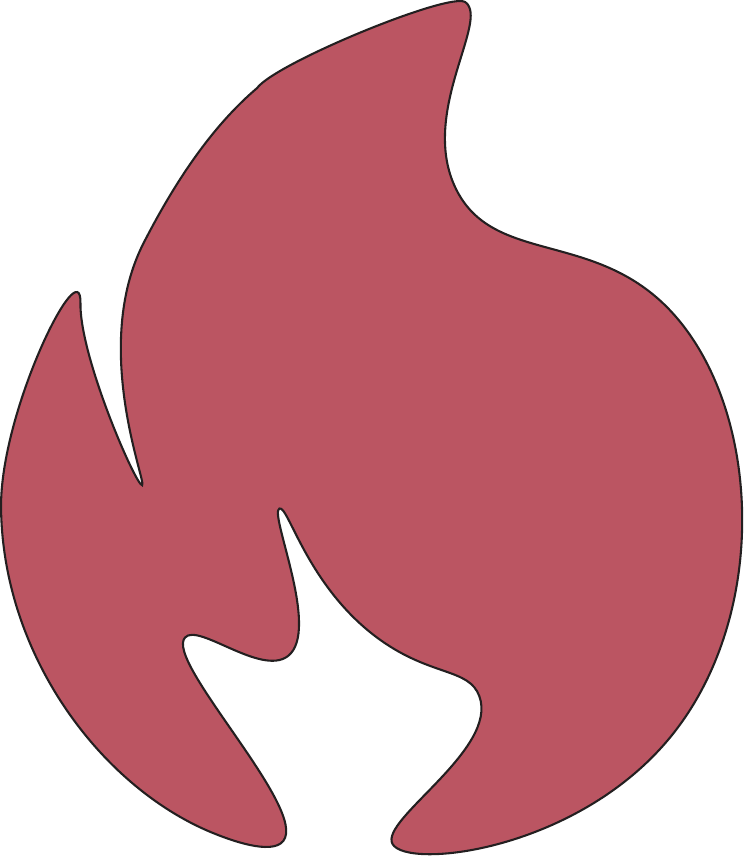} \inlinevis{-2pt}{1.1em}{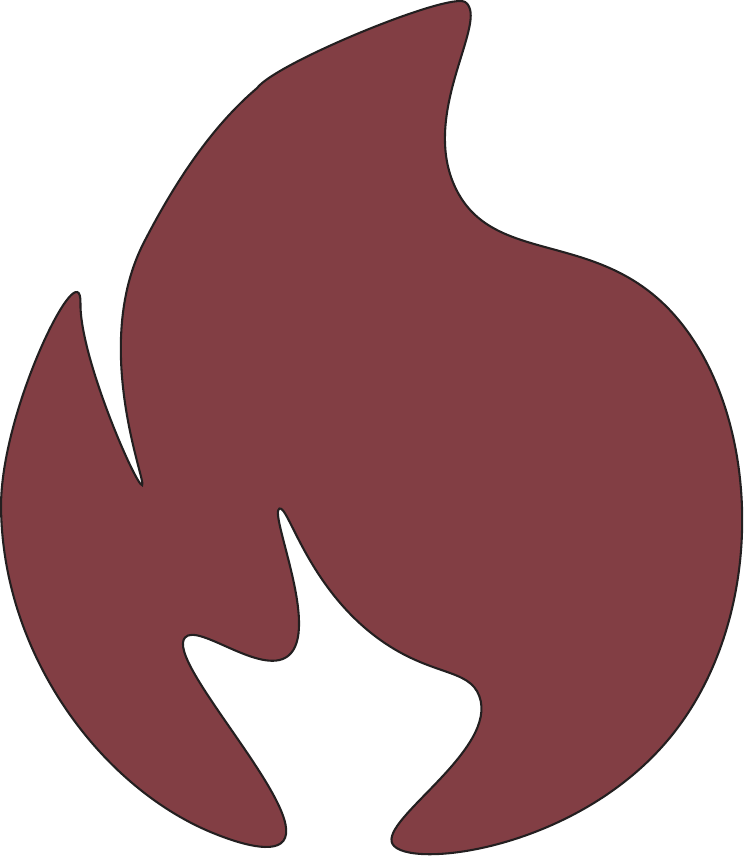}  \inlinevis{-2pt}{1.1em}{inline-icons/caloriesIcon}), size the icon relative to a quantity (\inlinevis{-2pt}{0.5em}{inline-icons/caloriesIcon-bright} \inlinevis{-2pt}{0.8em}{inline-icons/caloriesIcon-mediumdark}  \inlinevis{-2pt}{1.1em}{inline-icons/caloriesIcon}), or apply animation-related variables such as flicker. Icons can also be stacked to create unit-based pictographs. Similar examples we saw includ step counts represented by different shoe icons (e.\,g., walking, running, climbing), but examples were rare. %Not all quantities can be turned into categories for which meaningful icons exist. 

% In our previous survey, we created a table for each kind of data × representation type combination. However, the table has some gaps, especially in the only icon column. We filled these gaps by finding examples from premium watch faces on Facer and designing by ourselves. For the sleep stages, our visualization is meaningful only in specific scenarios. We display icons of different sleep stages on the watch, indicating the sleep stage that the user wakes up most often or a particular stage of sleep that the user lacks the most.
% We use flags or landmarks to visualize location names. This approach is only suitable for roughly expressing locations (divided by country only) or expressing well-known places.

\textbf{Charts} for data representation on watch faces were rare in our prior work and this review. However, charts have the potential to communicate information at a glance \cite{Tanja:2019:Glanceable-Visualization}, but need to be carefully designed to account for the small display size \cite{Healey:2012:Limits,Isenberg:2021:MicroVis}.

The \textit{chart types} that were commonly used to represent data on the premium smartwatch faces were bar charts, pie charts, donut charts, gauge charts, area charts, pictographs, and sliding scales. The data type suggests the most effective chart type to use as discussed above. The chart type directly affects its possible shape, size, and, ultimately also its location. For example, several types of charts can be curved and then aligned with the rim of a watch face. 
%The types of marks that make up a specific chart type also affect the colors that can be used as visualizations at small scale need often to be made high contrast to be easily readable. 
%And because of the different shapes of different charts, it indirectly affects the use of colors and the expression of themes.

The \textit{chart size} 
is related to the area occupied on the watch face. Size is more closely related to the shape of a chart rather than just its bounding box due to the tight layout of components on the watch. For bar charts and sliding scales, the size mainly refers to length and thickness. For pie, donut, and gauge charts, the size mainly refers to the diameter. For area charts and pictographs, the size mainly refers to their length and width. If a chart's shape matches a watch screen's shape, the chart can fill the watch face. For example, a donut chart can use the circumference of a round watch face. We rarely saw space-filling charts (similar to Dragicevic and Huot's SpiraClock \cite{Dragicevic:2022:SpiraClock}). Often charts were nested inside round containers similar to complications on physical pilot or diving watch faces. 

\textit{Chart color} describes how many hues are used on the chart, to style the chart or to represent data. Black and white watch face designs exist, in which hue cannot be used to represent data. In this case, textures may be used to represent categories \cite{zhong:hal-02944212}, but these are difficult to design well in practice. Decorators such as gradients affect a charts' color and have to be used with care. Gradients consist of different saturation or brightness of the same hue or gradients of different hues. One difficulty of using hues for categorical color encoding on watch faces is that the hues need to match the overall theme and aesthetic of the watch face. A color scheme that might create an aesthetic design overall might not work well for encoding data and vice versa. 

\textit{Chart position} 
refers to the place of the chart on the watch face. Often charts are positioned somewhere in the middle of the watch face or in the periphery. The shape of the chart affects the possible position of the chart. For round watches, ring-shaped or arc-shaped charts are most suitable for being located on the periphery of the watch face, whereas other shapes like bar- and line-charts can only be located in the middle of the watch face.

\textit{Chart theme} 
refers to the theme of complications, which often conforms to the overall topic of the watch face. Charts used in research rarely have a theme but for watch faces, these should be explored further. Colors, shapes, and in particular textures can help to define a theme. 

\revision{\subsection{Summary}
In summary, we present a design space for creating holistic watch faces considering four dimensions---externals such as personal and device factors, styles/themes, components, and representations---and their interplay in influencing smartwatch face design decisions. We further explore the various ways data can be represented on a watch face, including absolute numeric data, proportional data, categorical data, ordered data, temporal data, and geospatial data. Lastly, we delve into data representations on watch faces, such as text, icons, and charts, and discuss the challenges and possibilities of visualizing data on the limited display space of a smartwatch. In the next section, we discuss watch faces as small data dashboards and how our design space relates to dashboard design spaces.
}

\section{Discussion}

%In the previous section, we described a design space for smartwatch faces consisting of four dimensions---factors, styles/themes, components, and representations. We showed example values in these categories and how design decisions in these categories might influence each other. 
The goal of our design space is to describe existing watch faces \revision{and the factors that may influence visualization placement, design and reading}. These descriptions can be used to help compare different designs, to identify factors for assessing watch faces, and potentially help researchers generate new designs or variations of one design. In~\autoref{fig:design_space_examplary_watch_faces}, we show four descriptions of popular watch faces that have been generated with the dimensions of our design space. 
In this section, we discuss factors related to watch faces that go beyond our design space and shed more light on external factors and interactivity. 
\revision{We then deliberate on important and promising research directions in the next section.}

We consider watch faces as small data dashboards. Few~\cite{Few:2017:DashBoard} defines dashboards as ``a predominantly visual information display that people use to rapidly monitor current conditions that require a timely response to fulfill a specific role.'' Watch faces fit this definition well because they are visual displays that people glance at briefly to fulfill a specific information need. In their study on data dashboards Sarikaya et al. \cite{Sarikaya:2019:Dashboards} survey the purpose, audience, visual features, and data semantics. Likewise, Bach et al.~\cite{Bach:2022:Dashboard-Design} propose eight groups of design patterns that provide common solutions in dashboard design. 
Here, we make use of their dashboard design space and discuss how it relates to ours.

\bpstart{Purpose}
Sarikaya et al. \cite{Sarikaya:2019:Dashboards} describe two purposes for dashboards: decision support (strategic, tactical, operational) as well as communication and learning. Bach et al.~\cite{Bach:2022:Dashboard-Design} consider dashboards communication tools that focus on decision-making needs of: information overview, control, and conciseness. Our design space covers watch face purposes under external factors that influence the watch face design, its themes, components, and representation of complications. Of the two purposes Sarikaya et al.\ mention \cite{Sarikaya:2019:Dashboards}, watch faces also concern decision support (``do I need to walk more?'', ``should I leave for a meeting now?'') and communication in the wider sense by showing progress toward potential fitness goals, the weather, or simply showing the time. However, watch faces have other purposes that were not covered in this past work, for example, entertainment (some watch faces contain games with other complications) or simply making fashion statements. Further research is needed to elicit these purposes: how people choose watch faces, and how the purpose may guide design decisions in other parts of our design space. 

\bpstart{Audience}
According to Bach et al.~\cite{Bach:2022:Dashboard-Design}, novice dashboard users will need more instruction in the form of a clear layout and less information (e.g., single value, aggregated data), whereas expert or regular users will be more data literate and require more data and custom features on dashboards. Similarly, Sarikaya et al. \cite{Sarikaya:2019:Dashboards} argue that the visuals and functions of a dashboard ``reflect the intended audience, their domain and visualization experience, and their agency relationship with the data.'' Watch faces are ``individual'' dash boards in Saikaya et al.'s design space because they reside on a personal wearable device and are typically not shared with others. A broad range of the population buys smartwatches and there are now smartwatches targeted for children~\cite{zehrung2021investigating}. Designers of watch face visualizations, thus, cannot assume visualization literacy but many smartwatch wearers exhibit intricate familiarity with the data itself, their typical step counts, weather, or sleep patterns, for example. Most watch faces currently seem to be designed for people with low visualization literacy in that they mostly represent data through text and use simple graphs. Past work on word-scale visualizations \cite{Goffin:2017:AES,beck2017word}, however, has shown that complex representations are possible in a very small display space. 

\bpstart{Interactivity and Visual Features}
Interaction on watch faces has so far been quite restricted compared to what was found for more general dashboards \cite{Sarikaya:2019:Dashboards, Bach:2022:Dashboard-Design}. Interactions that we found relatively common were opportunities to tap through different types of data (e.\,g., switching from step counts to calories burned within one complication's container). We also found some interactive games that could be played by touching a watch face. Sometimes interactive changes to watch faces were enabled through companion apps, in which the layout and type of data shown on a watch face could be specified. What we have not seen is watch faces that allow to input data, correct data, or highlight specific complications. 
In terms of visual design, Bach et al.~\cite{Bach:2022:Dashboard-Design} emphasized color consistency in overall dashboard design and advocated for reducing color, if not required, particularly for screen fit dashboard designs.

\bpstart{Additional Data Semantics}
Several of Sarikaya et al.'s and Bach et al.'s additional data semantics apply to our work. In particular, ``benchmarks'' (e.\,g., step counts that have been reached) are commonly represented either by colors or by specific chart types, such as gauges and other proportion representations. Another similarity to their dashboards is regular data updates on watch faces and highlights when certain thresholds have been reached (e.\,g., critical battery levels or heart rates). These updates pose visualization challenges as outlined previously, especially for watch faces that are not monitored continuously.

%\revision{\subsection{Summary} The section above explored the design space for smartwatch faces and their relationship to data dashboards. 
%\bpstart{Summary} \revision{Watch faces serve as compact data dashboards, meeting specific information needs. Their purposes include decision support, communication, entertainment, and fashion statements, which require further investigation. We highlight that audiences with varying visualization literacy influence design choices, although more intricate representations are feasible in limited display spaces. Interactivity on watch faces is relatively restricted compared to general dashboards, offering options like swiping through data types and interactive games but lacking features for data input, correction, or highlighting complications. Last, prioritizing color consistency and visual design reduction is recommended, while additional data semantics like benchmarks and regular updates present visualization challenges for watch face dashboard designs.}

\section{Research Agenda}
While we have hinted at possible future research directions previously, here we highlight what we consider the most important or promising research directions for watch face visualization. 

%1) How much information can we perceive on a tiny screen?
\subsection{Holistic Study of Smartwatch Faces}
Watch face visualizations do not only need to be small but they also need to be integrated into a coherent watch face theme. As a personal data dashboard, watch faces not only make fashion statements but can also play an important role to facilitate decision making. We identified several open research questions that relate to the integration of visualizations in watch faces. 
%- How are they used? Do people see only tiny fractions? 
%- If they focus on sometime (e.g. time) how much other information to they see ``at a glance''

\bpstart{Designer's Vision vs.\ Use in Practice}
Through our prior work \cite{Islam:2020:Smartwatch-Survey} and current review, we observed a tension between a watch face design vision and the use of watch faces in practice. While a big part of the market designs included a single complication (18\%), 99\% of people in our prior work \cite{Islam:2020:Smartwatch-Survey} reported to have more than one complication on their watch face. We suspect that wearers often customize or personalize their watch faces to include more complications than the designer intended. Thus, while smartwatch designers may decide to follow design guidelines and recommended practices from the visualization community in their implementations, it is possible that smartwatch wearers make customization choices that violate these guidelines and practices. We can try to help wearers use a more appropriate design, for example, by suggesting best locations for placing a complication depending on clutter, background color, etc. Nevertheless, smartwatches remain personal devices, and wearers are (and should) ultimately be in control. As a community, we need to consider the right balance of allowing wearers to feel empowered in their customization, without being constrictive in our attempt to guide them.

%Smartwatch wearers should maintain their right to customize their devices (so preventing customization is not the solution). Rather, we should consider the bigger question about how to best educate wearers on visualization design. For example, we can envision extensions to the customization interface that provide guidance (e.\,g., suggest best locations for placing a complication depending on clutter, background color, etc.).

\bpstart{Study of the Influence of Themes}
The use of embellishments around data visualizations has been heavily debated in the visualization community. Some argue that they should be avoided \cite{tufteEnvisioning1990}; some argue that they do not hurt or can even help \cite{Bateman:2010:Chartjunk}. Watch face visualizations have, by design, a number of visual distractors surrounding them. UI styles such as skeuomorphism introduce extensive shading to give the illusion of depth, busy material textures and color palettes may distract from reading data. The influence of certain theme choices such as skeuomorphism vs.\ flat or semi-flat designs as well as the use of graphical decorators require further research. It is important to establish how they may support or hinder readability of data especially for watch face scenarios that require quick and accurate data readings.

\bpstart{Scalability: The Number of Complications and Their Complexity}
For small data representations, it is intuitive to recommend that their designs are simple and should encode only a few data values or dimensions. However, it would be useful to study the encoding limits empirically in more depth. Studies could investigate two scenarios of increasing visualization complexity: (1) adding more dimensions, more data values, more encoding types while keeping the same (small) display space and (2) decreasing the display space while keeping the same level of data complexity. Some of the past studies on visualization size (e.\,g.,~\cite{Cai:2018:Doughnut,Javed2010,Healey:2012:Limits,heer2009sizing,perin2013soccerstories}) have consistently shown that people prefer larger visualizations but we have also seen that participants effectively and correctly completed certain tasks with appropriate encodings at small scale \cite{Tanja:2019:Glanceable-Visualization}. In addition to questions regarding the scalability (or miniaturization) of visualizations, scalability questions also arise regarding the number of visualizations to show on watch faces. Should all data be represented with a single visualization? If not, what would be a good number to have? Both our prior survey and current review show that wearers had 3--5 complications on average, including time on their watch faces. The highest number of complications was 17. How many complications on a smartwatch display can effectively communicate with wearers needs further research. In summary, there are several avenues of scalability to explore: more data, smaller size, and more visualizations.

\bpstart{Specific Visualization Designs for Watch Faces}
Icons are useful on smartwatches as they often label the data shown. It would be useful to study how this label functionality of an icon can be combined with the display of numerical data. Possibilities exist such as using animation, generating pictographs, or sizing icons as discussed earlier but their effectiveness has not been empirically assessed for watch faces. Simplified icons with minimal color are highly recognizable \cite{Haroz:2015:ISOTYPE-Visualization} and potentially readable during quick glances providing motivation for studying them further. 

\vspace{-1mm}
\subsection{Smartwatch Technology and Interactivity}
In the future, we will see displays of different shapes emerge for which dedicated watch face designs have to be developed. First prototypes of watchstrap displays \cite{Klamka:2020:Watch+Strap} as well as curved displays \cite{Klamka:2021:Bendable-Color-EPaper-Displays} have emerged. Other types of embedded screens in clothing or as wristbands will emerge and we have to understand how watch faces can be designed for these non-flat displays. 

An issue related to display technology is how wearers can interact with the shown content. Few interactions with visualizations on watch faces have been implemented to mitigate ``fat finger problems'' (e.g., Bezel Interaction~\cite{Neshati:2021:BezelGlide}, EdgeSelect Interaction~\cite{Neshati:2022:EdgeSelect}). At this point mostly simple swipes and taps are used on touch-enabled watch faces. These allow, for example, switches between different representations or time intervals. Yet, more interactions that are complex have been explored for touch interactions on desktop-sized or tablet-sized charts (see \cite{lee2021post,brehmer2021interacting} for a summary). If they are useful and how they can be used in the context of smartwatches when the ``fat finger problem'' becomes even more acute is an interesting avenue for research. In addition to smartphones, many smartwatches are equipped with a microphone, allowing people to interact with them using speech. Therefore, recent research has started to investigate how to leverage touch and speech input for the collection or exploration of personal data on mobile devices~\cite{kim2021data,kim2022mymove,rey2023investigating}. It would be valuable to explore further, how to leverage speech input to facilitate more fluid watch face interactions. %Sometimes smart watches are not touch enabled, so input through speech or buttons also needs more research attention.

\subsection{The Role of Context}
%Adapted from petra's HDR
The primary intended context of a watch face's use needs to be considered in its graphical and interaction design. One important contextual factor is the potential movement of a wearer. Movement might lead to changing lighting conditions that can affect the readability of a watch face but movement also often entails a primary task such as driving or running. To what extent will movement affect people's ability to read a watch face such as during sports activities? The default for some Garmin watches, for example, is to show data during exercise using large black font on a white background without any visualizations. Is this the most effective and safe way to communicate data to wearers during other primary tasks?  Especially contexts with divided attention, for example, glancing while driving, require further research attention. While driving, viewers can only afford quick glances at watch faces. Visualizations in these settings are difficult to evaluate and test and future work is needed.

Another important factor is the intended task context for watch faces. \revision{While much research has centered around improving personal health through watch face representations, there needs to be more focus on the visualization challenges related to other specific tasks on smartwatch faces (see~\autoref{subsec:SmartwatchVisChallenges}).} We also saw commercial watch faces target contexts of use that we had not seen in research such as entertainment, festival, or military usage. Carpendale et al.~\cite{carpendale:2021:Mobile-Visualization-Design-Ideation} showed that with dedicated ideation exercises watch faces could be envisioned that target specific usage contexts such as sightseeing in their case. Digital watch faces are easy to switch and it would be interesting to study the impact of dedicated but changing watch faces on wearers. 

\section{Conclusion}
We presented a systematic review and design space for embedding visualizations into watch faces. The design space is grounded in the collection and analysis of properties of actual watch faces. Our holistic approach looked at smartwatches not just as data displays but also as a personal accessory for which visualizations are just one of many components that need to fit within a larger visual ecosystem. This led us to consider, for the first time, dimensions like the UI style and visual themes that may have an impact on visualization design, but have not been part of visualization design considerations in smartwatch research. We discuss the interplay of these dimensions and the choices available to visualization designers, as well as pitfalls and challenges when it comes to designing visualizations for such a personal use. Our research agenda highlights open opportunities both for visualization designers and for empirical research.

\ifCLASSOPTIONcompsoc
  \section*{Acknowledgments}
\else
  \section*{Acknowledgment}
\fi
In \autoref{fig:teaser} Smartwatch face examples come from Facer~\cite{Facer:2020}). From left to right: a) Pixel-like Analog / Digital (clayton), b) Big step 6000 (MB-Watch), c) B\# -- Open Heart (B Sharp Watches), d) Comic Pro! (Round or Square; Roch: Platinum Designs), e) Legion -- ctOS (Watch Dogs), f)~MOD--486 Weather Watch (Michael O'Day), g) Fall Mandala (Linlay Designs), h) Lovely (Cokanut).

The work was funded in part by ANR grant ANR-18-CE92-0059-01 and DFG grant DFG ER 272-14. Tanja Blascheck is funded by the European Social Fund and the Ministry of Science, Research, and Arts Baden-Württemberg.

\ifCLASSOPTIONcaptionsoff
  \newpage
\fi

\bibliographystyle{IEEEtran}
\bibliography{IEEEabrv,bibilography}

\begin{IEEEbiography}
    [{\includegraphics[width=1in,height=1.25in,clip,keepaspectratio]{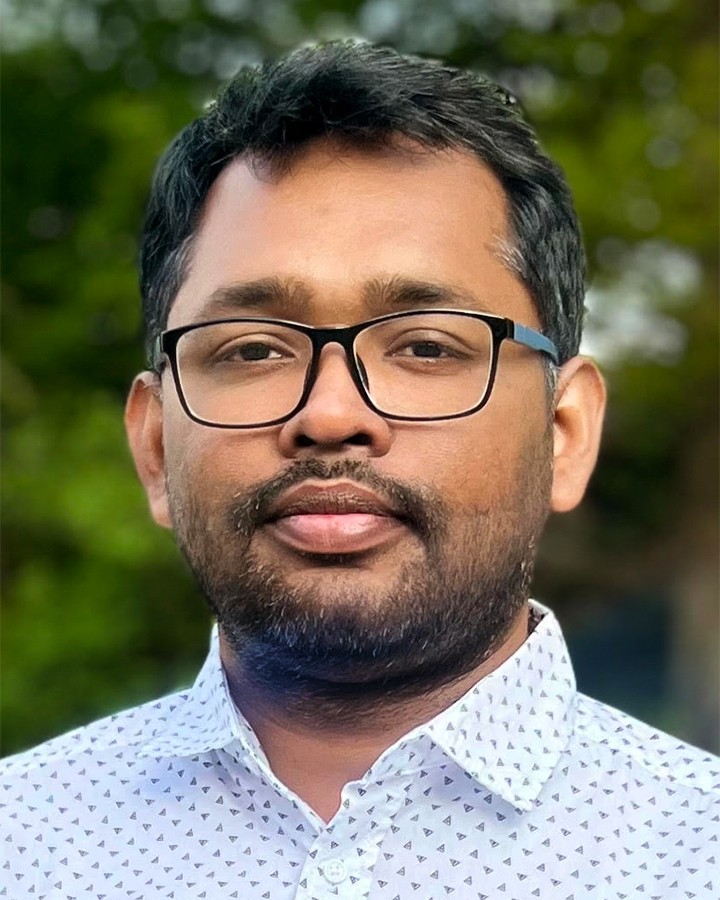}}]{Alaul Islam} is a postdoctoral fellow at the KITE Research Institute, University Health Network in Toronto, Canada. He recently completed his Ph.D. degree under the supervision of Dr. Petra Isenberg at Universit\'e Paris-Saclay and INRIA, France. His research interests focus on micro visualizations and health data visualization for wearables. 
\end{IEEEbiography}
\vskip 0pt plus -1fil

\begin{IEEEbiography}
    [{\includegraphics[width=1in,height=1.25in,clip,keepaspectratio]{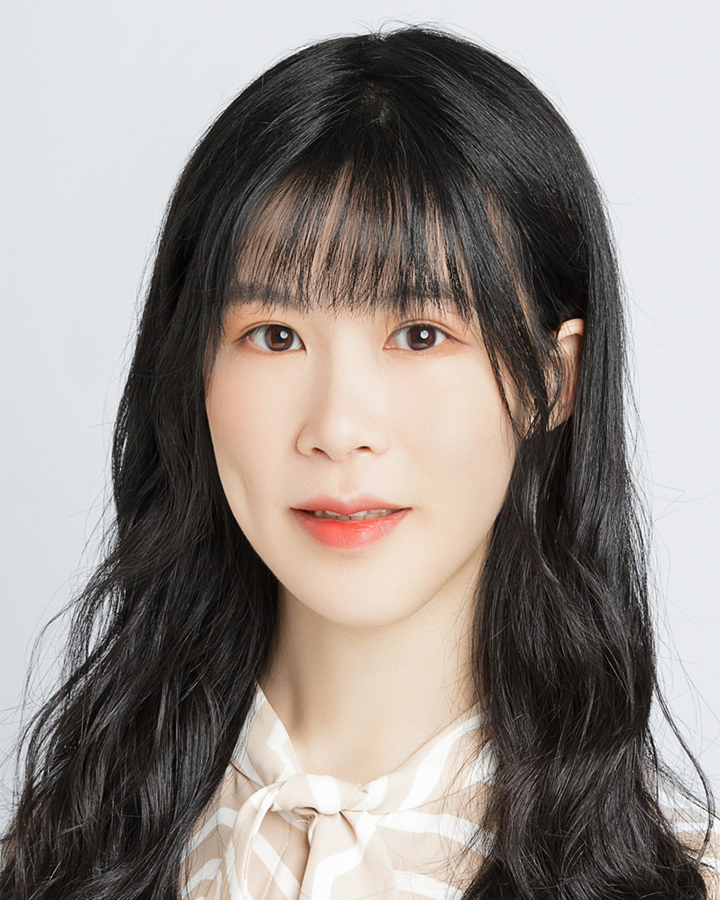}}]{Tingying He} is currently a second-year Ph.D. student working under the supervision of Dr. Tobias Isenberg at the Universit\'e Paris-Saclay, France. Her research is focused on exploring a design space visual data mapping for low-color displays. She also works on the evaluation of visualization and has developed a scale for measuring the aesthetic pleasure of visual representations.
\end{IEEEbiography}

\vskip 0pt plus -1fil
\begin{IEEEbiography}
    [{\includegraphics[width=1in,height=1.25in,clip,keepaspectratio]{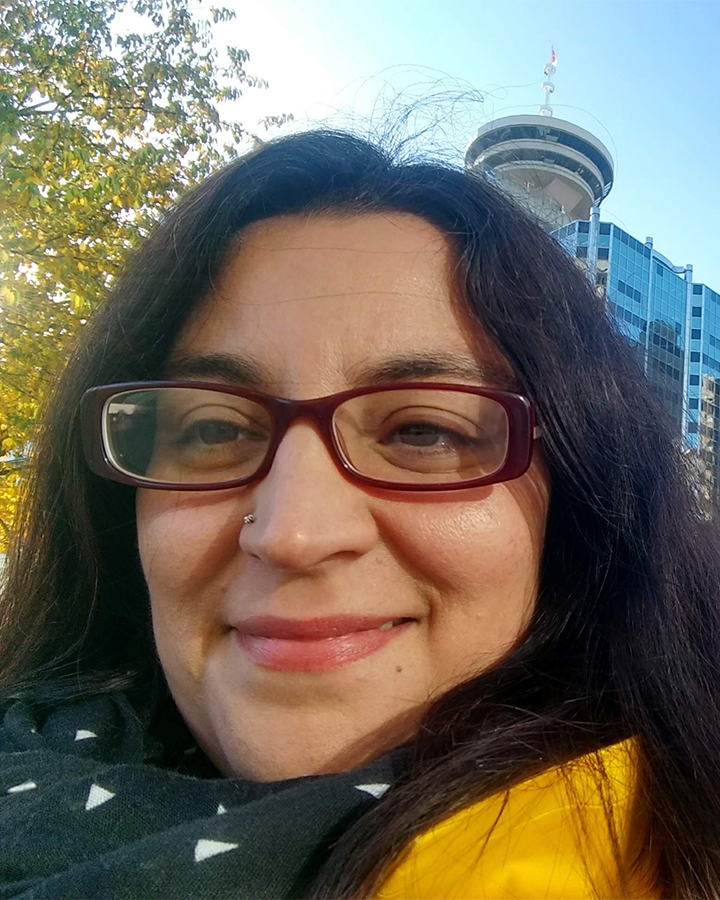}}]{Anastasia Bezerianos} is an Associate Professor with Universit\'e Paris-Saclay, Orsay, France and part of the ILDA Inria team. She works at the intersection of human-computer interaction and information visualization. Her research interests include  visualization evaluation, visual perception, decision making, and collaborative displays.
\end{IEEEbiography}
\vskip 0pt plus -1fil
\begin{IEEEbiography}
    [{\includegraphics[width=1in,height=1.25in,clip,keepaspectratio]{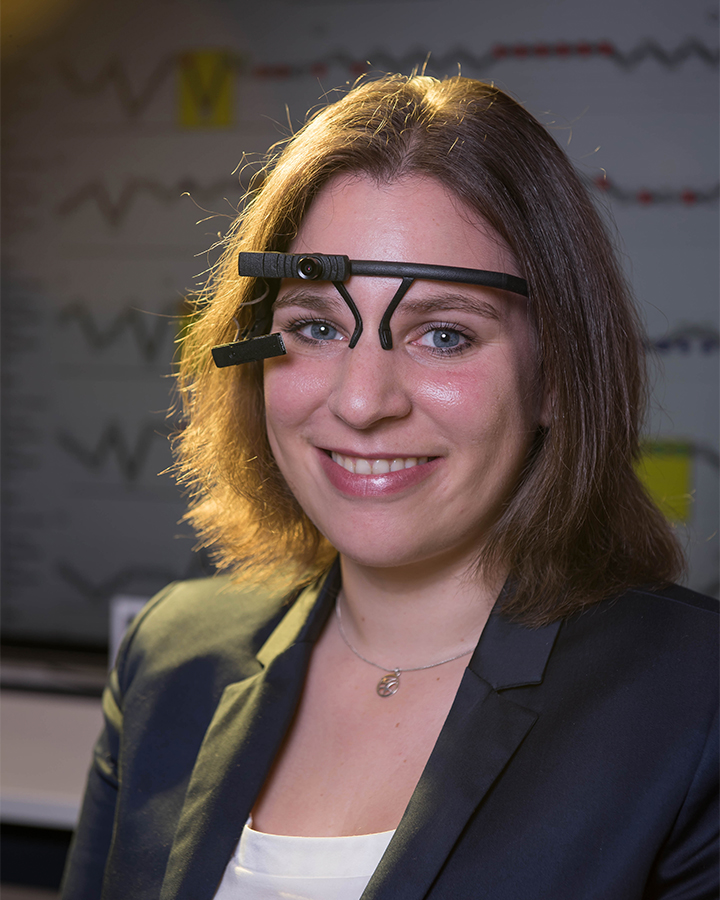}}]{Tanja Blascheck} is a Margarete-von-Wrangell Fellow and works at the Institute for Visualization and Interactive Systems at the University of Stuttgart. Her main research areas are information visualization and visual analytics with a focus on evaluation, eye tracking, and interaction.
\end{IEEEbiography}
\vskip 0pt plus -1fil
\begin{IEEEbiography}
    [{\includegraphics[width=1in,height=1.25in,clip,keepaspectratio]{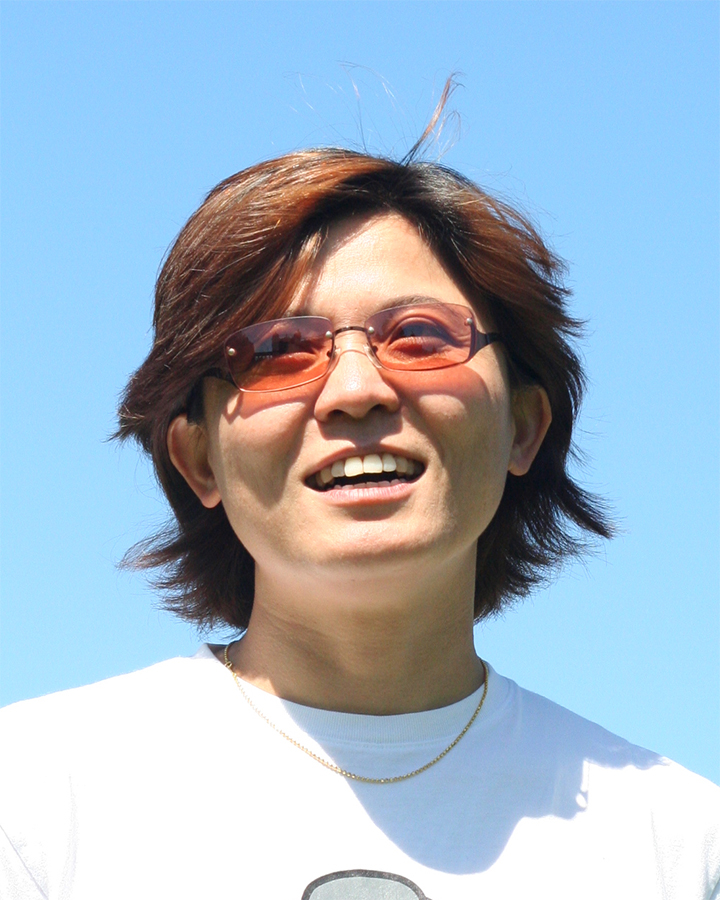}}]{Bongshin Lee} is a Senior Principal Researcher at Microsoft Research, where she conducts research on data visualization, human-computer interaction, and human-data interaction. With the overarching goal of empowering people with different abilities, she explores innovative ways to help them interact with data, by supporting easy and effective data collection, data exploration \& analysis, and data-driven communication. She received her Ph.D. in Computer Science from the University of Maryland at College Park.
\end{IEEEbiography}
\vskip 0pt plus -1fil
\begin{IEEEbiography}
    [{\includegraphics[width=1in,height=1.25in,clip,keepaspectratio]{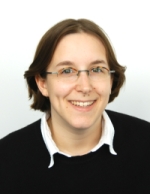}}]{Petra Isenberg} is a research scientist at Inria, France in the Aviz team. Her main research areas are visualization and visual analytics. She is interested in exploring how people can most effectively work when analyzing large and complex data sets---often on novel display technology such as small touch-screens, wall displays or tabletops.
\end{IEEEbiography}

\newpage
\begin{figure*}[t!]
\centering
\includegraphics[width=1\textwidth]{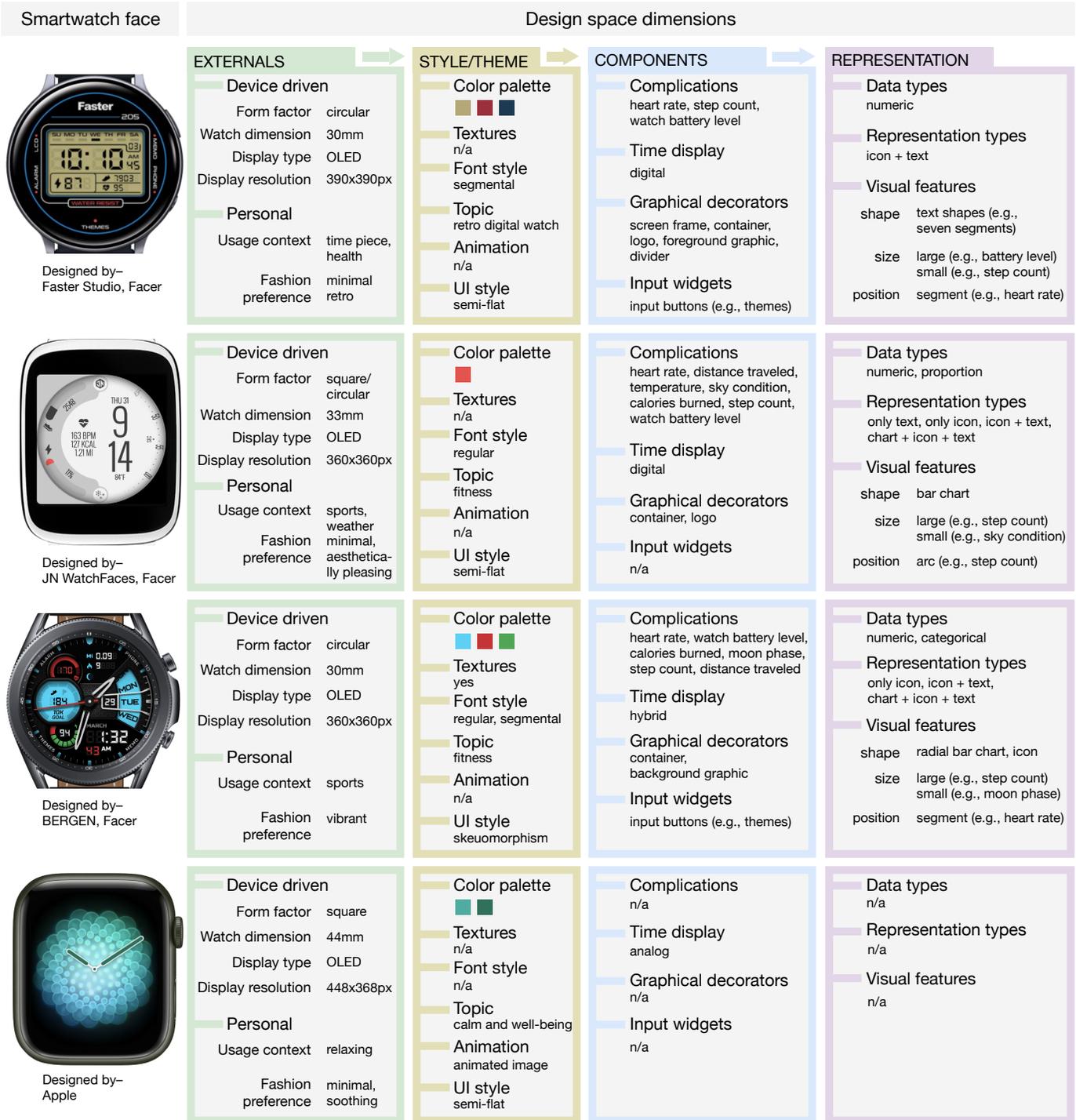}
  \caption{Four examples of smartwatch faces using our design space dimensions to describe them.
  }
%\vspace{-3mm}
  \label{fig:design_space_examplary_watch_faces}
\end{figure*}

\end{document}